\documentclass[a4paper]{article}
\usepackage{geometry}
\usepackage{amssymb}
\usepackage{bm}
\usepackage{amsthm}
\usepackage{mathtools}
\usepackage{natbib}
\usepackage{graphicx}
\usepackage[font=small]{caption,subcaption}
\usepackage{hyperref}
\usepackage{color}
\usepackage{float}    
\usepackage{authblk}
\usepackage[titletoc,title]{appendix}
\usepackage{setspace}
\usepackage{longtable}

\usepackage{bm}
\usepackage[titletoc,title]{appendix}
\usepackage{amsthm}
\usepackage{setspace}

\newtheorem{theorem}{Theorem}

\newtheorem{proposition}[theorem]{Proposition}

\newcommand{\bbeta}{\bm{\beta}}

\newcommand{\bZ}{\bm{Z}}

\newcommand{\bgamma}{\bm{\gamma}}
\newcommand{\btau}{\bm{\tau}}
\newcommand{\bzeta}{\bm{\zeta}}

\graphicspath{{figures/}}

\newcommand*{\thisdraft}{This version of draft: September 08, 2026}

\begin{document}

\title{
	Statistical inference for quantile correlation regression
	with bivariate outcomes  
	
}

\author[1]{Feipeng Zhang}  
\author[2]{Heng Peng
	\thanks{Correspondence Author: hpeng@hkbu.edu.hk}
}
\affil[1]{ 
	School of Economics and Finance, Xi'an Jiaotong University, Xi'an, 710049, China }
\affil[2]{
	Department of Mathematics, Hong Kong Baptist University, Hong Kong, China}

\date{\thisdraft}
\maketitle


\vspace*{-5mm}
\begin{abstract}
It is increasing scientifically important to recognize, quantify and visualize the dependence structure between the bivariate outcomes. 
In this paper, 
we investigate conditional dependence using a new  dependence measure, the \textit{quantile correlation coefficient}, 
which is defined by the Pearson correlation of two quantile-hits of bivariate outcomes. 
Furthermore, we propose a new regression model for covariate effects directly on the quantile correlation coefficient via a generalized linear regression model. 
The proposed model has intuitive interpretation and provides a comprehensive view of covariate effects across quantiles of bivariate outcomes. 
The proposed estimator enjoys desirable asymptotic properties including uniform consistency and weak convergence
as a process in the quantile index. 
Additionally,  
we develop the general hypothesis testing for the regression coefficients, with applications to summarize covariate effects across the quantile ranges. 
Simulation studies demonstrate the good finite sample performance of the proposed estimation and inference procedures. 
Two empirical applications of the proposed method 
reveal interesting insights in  premature twins data for measuring association and ChIP-seq data for measuring reproducibility.  
\end{abstract}

\noindent%
\vfill

\newpage
\setstretch{1.8}
\section{Introduction}
\label{sec:intro}

Bivariate outcome data, 
consisting of a pair of possibly correlated outcomes, 
are often encountered in  many applied sciences. 
For example, 
this data structure arises frequently in heritability studies, 
where the dependence between two random variables may shed insights into the genetic characteristics of traits or disease. 
In health economics, 
such data arises  when analyzing the gender gap in life expectancy
by examining the dependence between the life expectancy at birth of males  and females. 
Therefore, 
these practical scenarios require 
studying the dependence structure between two continuous random variables. 
Dependence measures have been proposed in previous literature to quantify the global strength of dependence, 
including Pearson coefficient, Kendall's tau, Spearman's rho, 
Gini’s coefficient,  concordance correlation coefficient, and the Blomqvist’s beta
~\citep{schweizer1981nonparametric,
	li1987concordance,
	hougaard2000analysis,josse2016measuring}. 
\cite{chatterjee2021new} recently proposed a new coefficient of correlation to measure degree of dependence. 
Another popular measure is to model the dependence structure between bivariate variables by a parametric copula with an unknown dependence parameter.  
The dependence parameter of a copula has a one-to-one relationship with the global dependence strength and can be estimated using likelihood-based or semi-parametric methods~\citep{nelsen2006introduction}.  
Various global dependence measures based on copulas have been developed, 
see e.g., 
\citet{zhang2019bet}, \cite{shih2021copula}, \cite{griessenberger2022multivariate} 
for recent surveys.   

Although the global dependence measures summarize the degree of dependence with a single readily interpretable number, 
it may be required to measure varying dependency between bivariate variables in many applications. 
For example, 
the correlation between financial
objects strengthens as the market declines, 
and approaches one when the market crashes, 
having the effect of destroying the benefit of diversification \citep{tjostheim2013local}.  
Therefore, it is desirable to consider the  local dependence measures that can characterize the dependency at any point of the region.  
Many local dependence measures has been proposed, 
including the
dependence function for densities~\citep{holland1987dependence}, 
local Kendall's tau~\citep{oakes1989bivariate}, 
local correlation extended Pearson coefficient
~\citep{bjerve1993correlation,doksum1994correlation,bairamov2003new}, 
local dependence function~\citep{jones1996local},  
local Gaussian correlation~\citep{tjostheim2013local}, 
and copula-based local correlations~\citep{ ledwina2015visualizing, shih2021copula}, 
among others.
For recent surveys,
see 
\cite{ding2017robust} and \cite{cmiel2020validation}.

In practice, 
it is often useful to evaluate covariate effects on the dependence between bivariate outcomes.
One motivating example comes from a retrospective study of premature twins, 
in which data on the birth weights of 200 twin pairs beyond 36 weeks of gestation were collected, along with their zygosity status~\citep{feng2009analysis}.
A natural question arising from this study is whether the strength of association in birth weights between twins is related to gestational age and zygosity. 
It may be expected that, conditional on gestational age, identical twins would exhibit greater similarity in birth weights than fraternal twins.  
Another motivating example is a ChIP-seq study of high-throughput biological experiments, 
where data were collected on two biological replicates and the binding sites being identified using the peak calling algorithms~\citep{chen2012systematic}. 
One primary scientific question is how to assess the impact of sequencing depth on the reproducibility of two biological replicates. 
This will provide insights into the depth required to achieve reliable results for different types of sites, 
thereby helping biologists identify the minimum sufficient depth that best fits their specific research goals. 
These two examples clearly motivate the interest in studying conditional dependence measures. 

To incorporate the additional covariate effects, 
there is a rich literature on conditional dependence measures, 
including 
kernel-based correlation coefficients
~\citep{huang2022kernel,sheng2023distance}, 
coefficients based on copulas
~\citep{gijbels2011conditional,dette2013copula,petersen2021testing,bianchi2023conditional}, 
conditional distance correlations
~\citep{
	wang2015conditional,fan2020projection}, 
generalized R-squared methods
~\citep{wang2017generalized,azadkia2021simple},
to name a few.  
However, the literature on  quantile-specific 
dependence measures, on the other hand, 
is not so large, especially in the local dependence setting.  
Undoubtedly, 
quantile-specific dependence measures may provide a clear view of the local dependence structure for the bivariate outcomes.  

The studies sharing  a similar concern with ours 
include 
the quantile association regression based on quantile-specific odds ratio~\citep{li2014quantile}, 
quantilogram and cross-quantilogram~\citep{linton2007quantilogram,han2016cross,lee2020quantilograms},
and 
quantile correlation~\citep{li2015quantile}.  
Specifically, \cite{li2014quantile} proposed regression models for covariate effects on quantile association measures, 
the quantile-specific odds ratio and the quantile-specific probability ratio. 
However, 
their dependence measures have no intuitive interpretation for practitioners who are not familiar with odds or probability ratios.  
\cite{han2016cross} developed theoretical results for the cross-quantilogram \citep{linton2007quantilogram} to measure the quantile dependence between two time series, 
which is exactly the correlation of their quantile-hits. 
\cite{lee2020quantilograms} further  investigated the quantilogram and cross-quantilogram under the presence of long-range dependence. 
While it is conceptually appealing and simple to interpret, 
quantilogram is only a measure of serial dependence between  two time series.   
Alternatively, \cite{li2015quantile} defined a quantile correlation by the correlation of one random variable and the quantile-hit of another random variable. 
Since it is not symmetric, this measure has no clear interpretation related to tail dependence and cannot be compared across different quantile levels within a given distribution.  


In this paper,  
we define a new local dependence measure indexed by the quantiles of bivariate outcomes that captures the degree of tail dependent association well and allows for comparisons across different quantile levels within a given distribution. 
Our first contribution 
is to propose a new coefficient of local dependence between two random variables,  
named as the \textit{quantile correlation coefficient}. 
Inspired by quantilogram proposed by \cite{han2016cross}, 
the quantile correlation coefficient
is the Pearson correlation coefficient of two quantile-hits of bivariate outcomes. 
The main appealing features of our coefficient are the following: 
(1) it has a simple expression with range [-1,1],
(2) it inherits the merits of the Pearson correlation coefficient, but can measure the nonlinear dependence between 
bivariate outcomes,
(3) it is robust for bivariate outcomes with heavy-tailed
distributions,
(4) it is independent of the marginal distributions,  
and is invariant under monotone transformations. 

The second contribution of this paper is 
to develop a new regression model for covariate effects on the quantile correlation coefficient. 
Specifically,  
we formulate the dependence regression model on the quantile correlation coefficient via a generalized linear regression model, 
which provides a comprehensive view of covariate effects across different quantile levels of bivariate outcomes.  
The regression parameters are estimated by constructing an unbiased estimating equation. 
We establish asymptotic properties of the proposed estimator, 
including uniform consistency and weak convergence as a process of the quantile index. 
Furthermore, 
we develop the general hypothesis test for the regression coefficients, with applications to test for summarizing covariate effects across quantile ranges. 
Additionally, we propose to calculate critical values by the Gaussian multiplier bootstrap, 
which resamples the linear approximations of the empirical inference processes.

The remainder of the paper is organized as follows. 
Section~\ref{s:method} introduces the statistical model and method, 
including the definition and properties of the quantile correlation coefficient,  and the quantile correlation regression framework. 
The  unbiased estimating equation estimator of regression parameters and its asymptotic properties are also presented in this section. 
Section~\ref{s:test} develops inference procedures for the general hypothesis testing for the regression coefficients, and presents practical inference for testing distributional hypotheses. 
In Section~\ref{s:sim}, we conduct extensive Monte Carlo simulations to evaluate the finite sample properties of the proposed estimation and inference procedures. 
Section~\ref{s:app} presents two empirical illustrations,   
prematrue twins data for measuring birth weights' association  
and  ChIP-seq data for measuring reproducibility of high-throughput experiments. 
Section~\ref{s:con} concludes the paper with some discussions. 
Supplementary materials are devoted to technical proofs and addition numerical results.

\section{Statistical Model and Method}
\label{s:method}

\subsection{Quantile correlation coefficient}
\label{ss:qcc}

In this section, we propose a  new local dependence measure indexed by the quantiles of the bivariate outcomes, 
named as the \textit{quantile correlation coefficient}, 
which is the Pearson correlation coefficient of two quantile-hits of bivariate outcomes. 

For a pair of continuous outcomes $(Y_1,Y_2)$, 
we define the quantile correlation coefficient (qcc) 
at a pre-specified $\btau=(\tau_1,\tau_2)\in \mathbb{D}$ 
by 
\begin{equation}
	\label{eqn:qcc}
	\begin{aligned}	
		\operatorname{qcc}_{(Y_1,Y_2)}(\btau)
		&=
		\operatorname{cor}\left(
		I\left\{Y_1\leq Q_{\tau_1}(Y_1)\right\}, 
		I\left\{Y_2\leq Q_{\tau_2}(Y_2)\right\}
		\right)\\
		&=
		\frac{\operatorname{E}\left[
			\psi_{\tau_1} \left\{Y_1-Q_{\tau_1}(Y_1)\right\}
			\psi_{\tau_2} \left\{Y_2-Q_{\tau_2}(Y_2)\right\}
			\right]}
		{\sqrt{\operatorname{E}\left[
				\psi_{\tau_1}^2 \left\{Y_1-Q_{\tau_1}(Y_1)\right\}
				\right]}
			\sqrt{\operatorname{E}\left[
				\psi_{\tau_2}^2 \left\{Y_2-Q_{\tau_2}(Y_2)\right\}
				\right]}
		},
	\end{aligned} 
\end{equation}
where $Q_{\tau_j}(Y_j)$ is the $\tau_j$-th quantile of $Y_j$ for $j=1,2$, 
$\mathbb{D}$ is a compact subset of the unit square $(0,1)^2$,  
$\psi_\tau(u)=\tau- I(u\leq 0)$ is the subgradient of the check loss function $\rho_\tau(u)=u[\tau- I(u\leq 0)]$, and $I(\cdot)$ is the indicator function.

Basically,  qcc is the Pearson correlation coefficient of two quantile-hits $I\left\{Y_1\leq Q_{\tau_1}(Y_1)\right\}$  and 	
$I\left\{Y_2\leq Q_{\tau_2}(Y_2)\right\}$,
which are the pair $(Y_1,Y_2)$ simultaneously less than their $\tau_1$th and  $\tau_2$th quantiles respectively.  
Thus, its range is between -1 and 1, 
and especially  qcc equals to 0 if $Y_1$ and $Y_2$ are independent. 
However, in contract to the usual Pearson coefficient only measuring the linear dependence, 
qcc can measure the nonlinear dependence between $Y_1$ and $Y_2$. 
Besides,  qcc is robust for both $Y_1$ and $Y_2$ with heavy-tailed distributions, because the quantile-hit as the indicator function is not affected by outliers and extreme values. 
More importantly, 
qcc parameterizes the dependence in the simple form of the Pearson correlation coefficient for the binary outcomes whether the component outcome is less than its quantile. 
Thus, it assesses the dependence locally, reflecting the tendency of the pairs $(Y_1,Y_2)$ to stay on the same side of their respective $\tau_1$th and $\tau_2$th quantiles. 

In what follows, 
we write $\operatorname{qcc}(\btau)$
as $\operatorname{qcc}_{(Y_1,Y_2)}(\btau)$  for brevity. 
The following proposition summarizes the main properties of qcc. 
\begin{proposition}
	\label{prop:qcc}
	$\operatorname{qcc}_{(Y_1,Y_2)}(\btau)$ of $(Y_1,Y_2)$ has the following main properties: 
	\begin{itemize}
		\item [(i)]
		$-1\leq \operatorname{qcc}_{(Y_1,Y_2)}(\btau) \leq 1$ for all $\btau\in \mathbb{D}$. 
		
		\item [(ii)]
		$\operatorname{qcc}_{(Y_1,Y_2)}(\btau)=0$ if and only if $Y_1$ and $Y_2$ are independent.  
		
		\item [(iii)]
		$\operatorname{qcc}_{(Y_1,Y_2)}(\btau)$ is maximal (minimal) if and only if $Y_2=g(Y_1)$ and $g$ is strictly increasing (decreasing) a.s. on the range of $Y_1$.

		\item [(iv)] 
		For any strictly increasing functions $g_1$ and $g_2$, 
		$\operatorname{qcc}_{(g_1(Y_1), g_2(Y_2))}(\btau)=\operatorname{qcc}_{(Y_1,Y_2)}(\btau)$. 
	\end{itemize}
\end{proposition}


\textbf{Remarks~1}. 
\begin{itemize}
	\item[(1).]
	$\operatorname{qcc}$ inherits the merits of the Pearson correlation coefficient. Essentially, however, 
	it can measure the nonlinear dependence between bivariate outcomes $(Y_1,Y_2)$, and hence it can incorporate different kinds of bivariate joint distributions of $(Y_1,Y_2)$. 
	
	\item[(2).]
	By its definition, qcc can be simply expressed by 
	$
	\operatorname{qcc}(\btau)=\frac{F\left\{Q_{\tau_1}(Y_1), Q_{\tau_2}(Y_2)\right\}-\tau_1\tau_2}{\sqrt{(\tau_1-\tau_1^2)(\tau_2-\tau_2^2)}  },
	$
	where $F(\cdot,\cdot)$ is the joint distribution of $(Y_1,Y_2)$. 
	Obviously, 
	qcc can be viewed as a direct connection with the joint distribution. 
	Then, qcc can be also related to the copula-based partial correlation in \citet{xia2021copula}. 
	Formally, the formula of $\operatorname{qcc}$ is exactly the copula-based measure of dependence, quantile dependence function proposed by \cite{cmiel2020validation}, see more basic properties therein. 
	However, our proposed quantile correlation coefficient does not depend on copula framework, and hence it is very general and flexible for measuring dependence of bivariate outcomes. 
	
	\item[(3).]
	For $n$ independent and identically distributed (IID) replicates $\{(Y_{i1}, Y_{i2}): i=1,\ldots,n\}$ 
	from $(Y_1, Y_2)$, 
	a natural estimator of $\operatorname{qcc}(\btau)$ is
	the sample version of qcc,  
	\begin{equation}
		\label{eqn:qcc-sample}
		\begin{aligned}	
			\widehat{\operatorname{qcc}}(\btau)
			=
			\frac{
				n^{-1}\sum\limits_{i=1}^n
				I\left\{Y_{i1}\leq \widehat{Q}_{\tau_1}(Y_1), 
				Y_{i2}\leq \widehat{Q}_{\tau_2}(Y_2)\right\}-\tau_1\tau_2}{\sqrt{(\tau_1-\tau_1^2)(\tau_2-\tau_2^2)}  },
		\end{aligned} 
	\end{equation}
	where $\widehat{Q}_{\tau_j}(Y_j)$ is the sample $\tau_j$-th quantile of $Y_j$ for $j=1,2$. 
\end{itemize}

\subsection{Graphical illustration}
\label{ss:graphical}

\begin{figure}
	\centering
	\begin{tabular}{ccc}
		\makebox[0pt][r]{\makebox[20pt]{\raisebox{40pt}{\rotatebox[origin=l]{90}{Clayton}}}}%
		\includegraphics[width=0.32\textwidth]{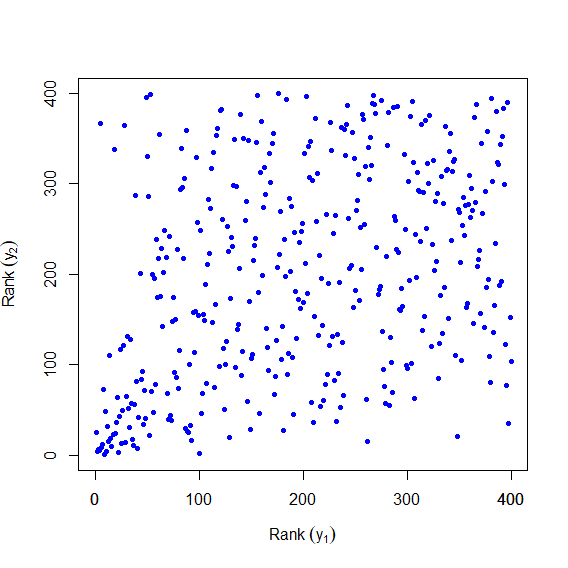} & \includegraphics[width=0.32\textwidth]{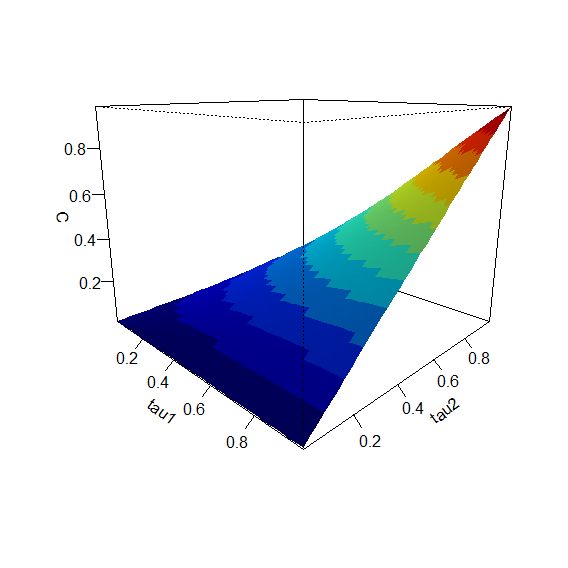} &
		\includegraphics[width=0.32\textwidth]{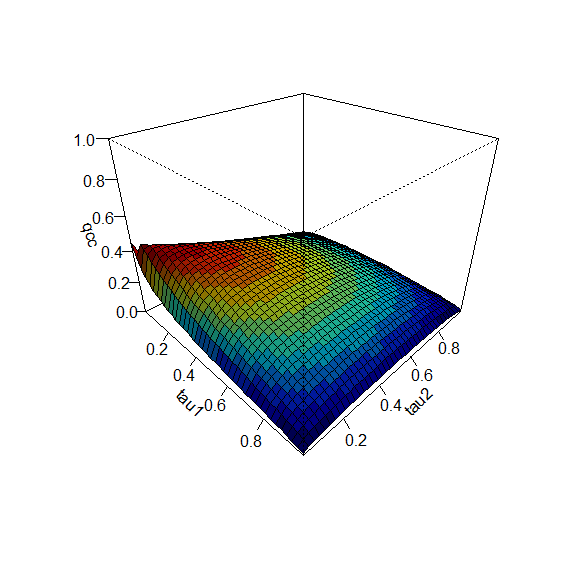} \\
			\vspace*{-2mm}
		\makebox[0pt][r]{\makebox[20pt]{\raisebox{40pt}{\rotatebox[origin=l]{90}{Gumbel}}}}%
		\includegraphics[width=0.32\textwidth]{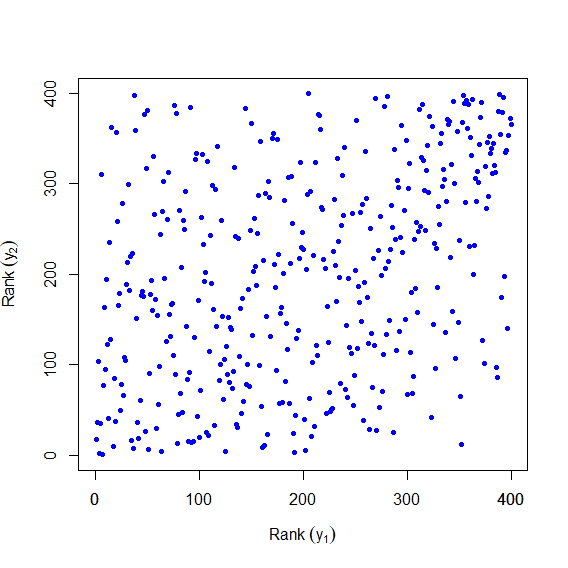} & \includegraphics[width=0.32\textwidth]{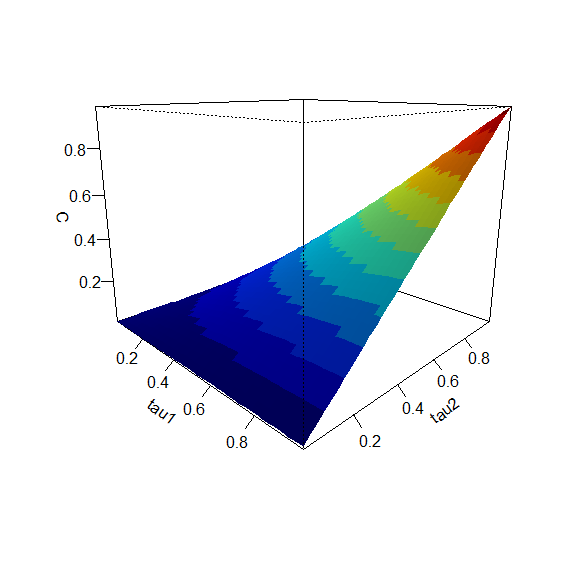} &
		\includegraphics[width=0.32\textwidth]{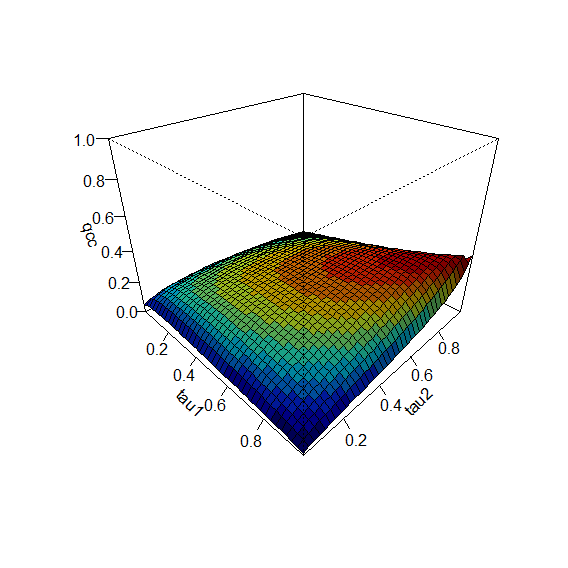} \\
			\vspace*{-2mm}
		\makebox[0pt][r]{\makebox[20pt]{\raisebox{40pt}{\rotatebox[origin=l]{90}{Normal}}}}%
		\includegraphics[width=0.32\textwidth]{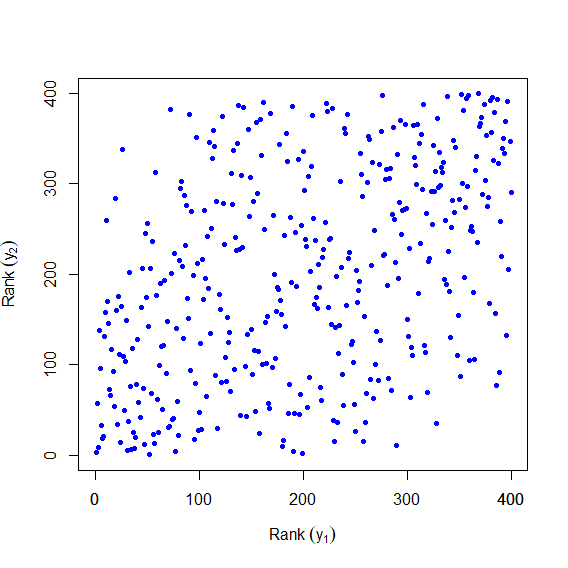} & \includegraphics[width=0.32\textwidth]{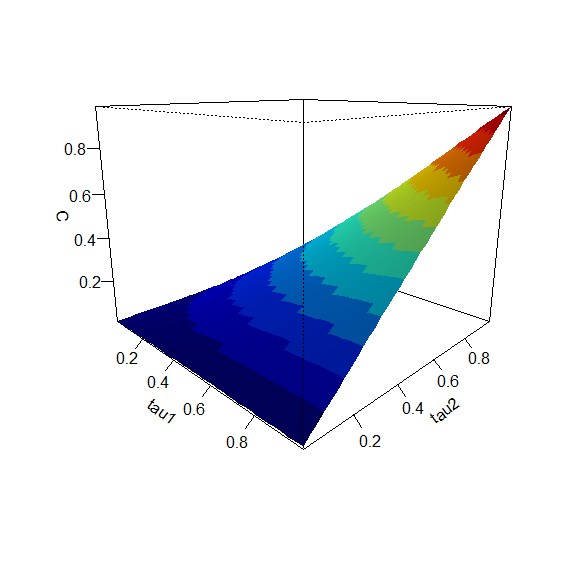} &
		\includegraphics[width=0.32\textwidth]{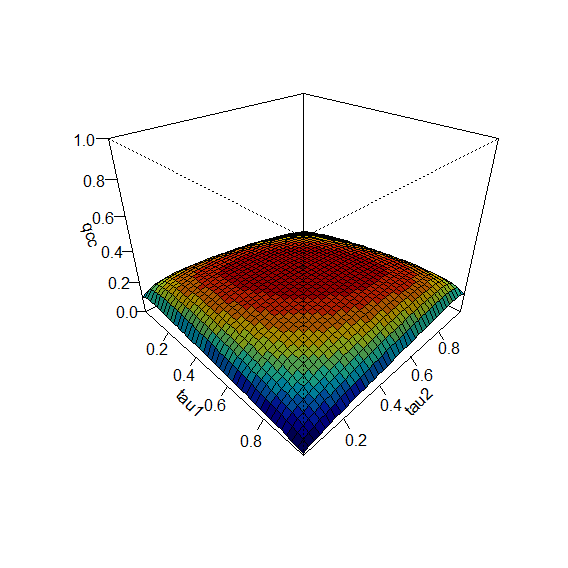} \\
			\vspace*{-2mm}
			\makebox[0pt][r]{\makebox[20pt]{\raisebox{40pt}{\rotatebox[origin=l]{90}{FGM}}}}%
		\includegraphics[width=0.32\textwidth]{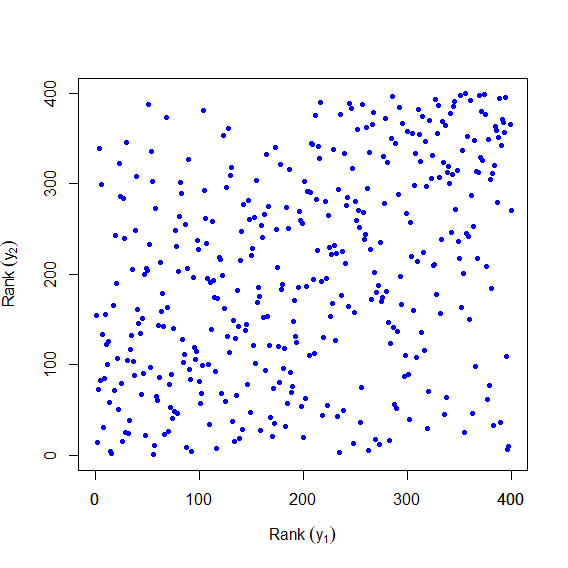} & \includegraphics[width=0.32\textwidth]{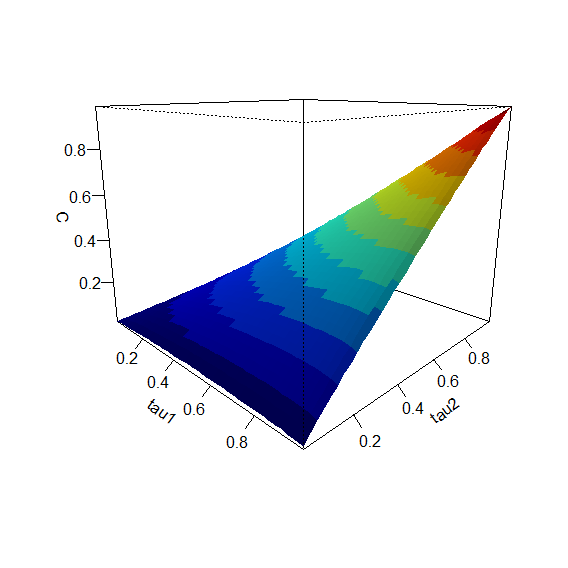} &
		\includegraphics[width=0.32\textwidth]{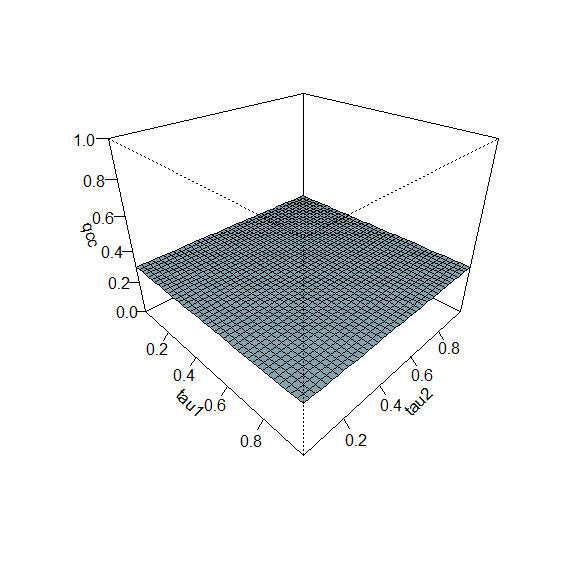} \\
	\vspace*{2mm}
		{\small (a) Rank scatterplot} & 
		{\small (b) $C(\tau_1,\tau_2)$} & 
		{\small (c) $\operatorname{qcc}(\tau_1,\tau_2)$} \\
	\end{tabular}
		\caption{Heuristic examples for comparison of the rank scatterplot, copula function $C(\tau_1,\tau_2)$ and  quantile correlation coefficient $\operatorname{qcc}(\tau_1, \tau_2)$
		under different copulas. From top to bottom, the copulas are Clayton, Gumbel, Normal and FGM, respectively. 
	}
	\label{fig:baseline}
\end{figure}

To visually understand qcc, 
we compare the shapes of rank scatterplot, 
copula function $C(\tau_1,\tau_2)$ and  
quantile correlation coefficient $\operatorname{qcc}(\tau_1, \tau_2)$
when the underlying copulas are 
Clayton, Gumber, Normal and FGM 
in Figure~\ref{fig:baseline}. 
The copula parameters are chosen by satisfying the global Kendall's tau as 0.3.  
Graphically, the overall dependence is similar in copula function, 
but $\operatorname{qcc}(\tau_1, \tau_2)$ can successfully capture 
the big differences in local dependence levels.  
Compared to the shape of $C(\tau_1,\tau_2)$,  
$\operatorname{qcc}(\tau_1, \tau_2)$ can provide more straightforward insights into the dependence structure.  
Specifically, 
$\operatorname{qcc}(\tau_1, \tau_2)$ varies with $(\tau_1,\tau_2)$,  
whose magnitude and shape reflect both the dependence strength and structure, 
notably, it remains constant across $(\tau_1,\tau_2)$ only for the FGM copula. 
Besides, $\operatorname{qcc}(\tau_1, \tau_2)$ can capture the lower, upper tail and symmetry of Clayton, Gumbel and Normal copulas, respectively. 
Thus, in exploratory analysis, 
it is of great interest to investigate 
$\operatorname{qcc}(\tau_1, \tau_2)$ 
in a wide range of $(\tau_1,\tau_2)$ to gain a complete understanding the dependence structure.

\begin{figure}
	\centering
	\begin{subfigure}{0.45\textwidth}	
		\includegraphics[width=3in]{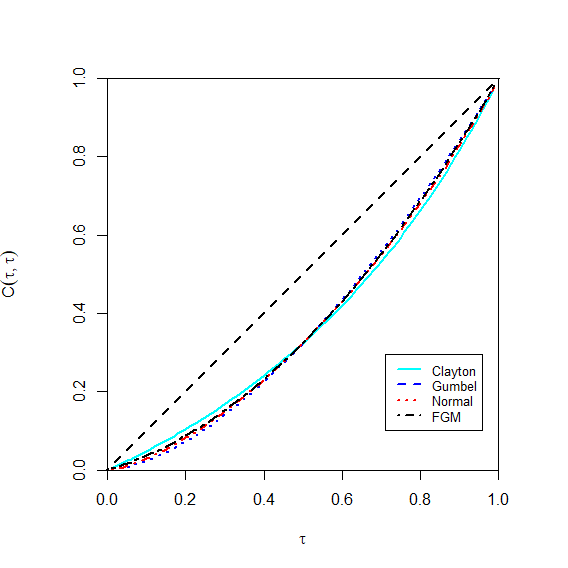}
		\caption{$C(\tau,\tau)$}
	\end{subfigure}	
	\hspace{2em}
	\begin{subfigure}{0.45\textwidth}	
		\includegraphics[width=3in]{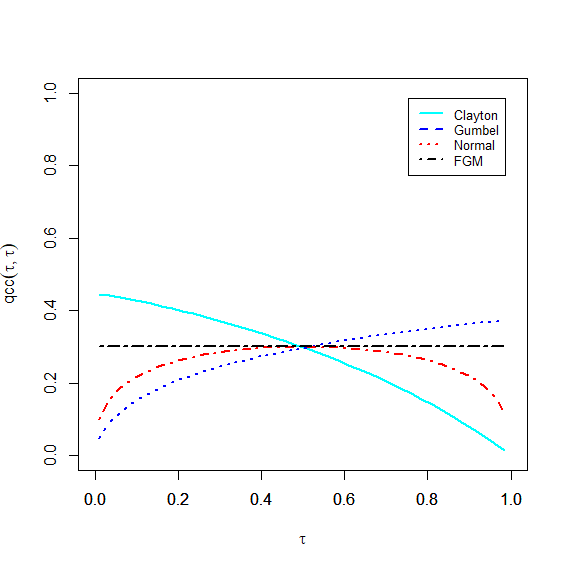}
		\caption{$\operatorname{qcc}(\tau,\tau)$}
	\end{subfigure}	
	\caption{Shapes of $C(\tau,\tau)$ and $\operatorname{qcc}(\tau,\tau)$ for Clayton, Gumbel, Normal and FGM copulas.}
	\label{fig:shape}
\end{figure}

In many applications, 
it may be particularly interested in  $\operatorname{qcc}(\tau_1, \tau_2)$ at certain $(\tau_1,\tau_2)$, for example, the diagonal line $\tau_1=\tau_2=\tau$ in the birth weights of premature twins. 
This line is of particular interest when the bivariate outcomes are positively correlated. 
Then, $\operatorname{qcc}(\tau, \tau)$ admits an intuitive interpretation and measures the concordant tendency of the bivariate outcoms below their $\tau$-th quantiles. 
Figure~\ref{fig:shape} depicts  the shapes of 
$C(\tau_1,\tau_2)$ and 
$\operatorname{qcc}(\tau, \tau)$ for Clayton, Gumbel, Normal and FGM copulas. 
The shape of $C(\tau,\tau)$ in Figure~\ref{fig:shape}(a) can not be  visually distinguishable for these four different copulas,   
but the shape of $\operatorname{qcc}(\tau, \tau)$ in Figure~\ref{fig:shape}(b) provides an accurate description of the local dependence level and trend. 
Clearly, $\operatorname{qcc}(\tau, \tau)$ is high at small quantiles for Clayton copula, while is high at large quantiles for Gumbel copula. 
$\operatorname{qcc}(\tau, \tau)$ of Normal copula exhibits symmetric pattern against $\tau=0.5$, 
and it remains constant across $\tau$ for FGM copula.

\subsection{Quantile correlation regression}
\label{ss:qcc-reg}

Most importantly, qcc at any  quantile level is simply an Pearson correlation coefficient, which is standard to assess the dependence of bivariate outcomes. 
This motivates us to directly formulate a dependence regression model on $\operatorname{qcc}(\btau)$, rather than on $C(\btau)$ for $\btau=(\tau_1,\tau_2)\in \mathbb{D}$, 
could render a more sensible method to model covariate effects on local dependence.  
To incorporate the covariate effects $\bZ$, we can define the conditional quantile correlation coefficient at a pre-specified $\btau=(\tau_1,\tau_2)^\top$ as 
\begin{equation}
	\label{eqn:cqcc}
	\begin{aligned}	
		\operatorname{qcc}(\btau|\bZ) 
		&= 
		\operatorname{cor}\left(I\left\{Y_1\leq Q_{\tau_1}(Y_1|\bZ)\right\}, 
		I\left\{Y_2\leq Q_{\tau_2}(Y_2|\bZ)\right\}
		|\bZ
		\right)\\
		&=
		\frac{\operatorname{E}\left[
			\psi_{\tau_1} \left\{Y_1-Q_{\tau_1}(Y_1|\bZ)\right\}
			\psi_{\tau_2} \left\{Y_2-Q_{\tau_2}(Y_2|\bZ)\right\}
			|\bZ
			\right]}
		{\sqrt{\operatorname{E}\left[
				\psi_{\tau_1}^2 \left\{Y_1-Q_{\tau_1}(Y_1|\bZ)\right\}|\bZ
				\right]}
			\sqrt{\operatorname{E}\left[
				\psi_{\tau_2}^2 \left\{Y_2-Q_{\tau_2}(Y_2|\bZ)\right\}|\bZ
				\right]}
		},
	\end{aligned}
\end{equation}
where $\bZ=(1, Z_1,\ldots, Z_p)^\top$ is a $(p+1)$-dimensional covariate vector, 
and for $j=1, 2$, 
$Q_{\tau_j}(Y_j|\bZ)= \inf\left\{y: F_j(y|\bZ)\geq \tau_j \right\}$
is the marginal quantiles of $Y_j$ conditional on $\bZ$, 
and 
$F_j(y|\bZ)=P(Y_j\leq y|\bZ)$ is the conditional 
marginal distribution of $Y_j$ given $\bZ$.  
Some algebra can show that  $qcc(\btau)$ can be equivalent to 
\begin{equation}
	\label{eqn:cqcc-equiv}
	\operatorname{qcc}(\btau|\bZ) =
	\frac{F\left\{Q_{\tau_1}(Y_1|\bZ), Q_{\tau_2}(Y_2|\bZ)|\bZ\right\}-\tau_1\tau_2}{\sqrt{(\tau_1-\tau_1^2)(\tau_2-\tau_2^2)}  },
\end{equation}
where $F(\cdot,\cdot|\bZ)$ is the joint distribution of $(Y_1,Y_2)$ given $\bZ$. 
In view of \eqref{eqn:cqcc-equiv}, $\operatorname{qcc}(\btau|\bZ)$ 
clearly depends on 
the conditional joint distribution $F(\cdot,\cdot|\bZ)$ and two marginal conditional quantiles $Q_{\tau_1}(Y_1|\bZ)$
and  $Q_{\tau_2}(Y_2|\bZ)$. 
For simplicity and without loss of generality, 
we can model the marginal $Q_{\tau_j}(Y_j|\bZ)$ for $j=1,2$, 
by the standard linear quantile regression,
\begin{align}
	\label{eqn:mod-marginal}
	Q_{\tau_j}(Y_j|\bZ)=\bZ_j^\top \bbeta_j(\tau_j),  \quad \tau_j\in (0,1), 
\end{align}
where $\bZ_j$ is $(p_j+1)\times 1$ subvector of $\bZ$ that may overlap each other and includes 1 as the first element, 
and the $(p_j+1)\times 1$ regression coefficient $\bbeta_j(\tau_j)$ is an  unknown function of $\tau_j$. 
Hence, 
$\bbeta_j(\tau_j)$ can be estimated by the standard linear quantile regression. 
On the other hand, 
for  $F(\cdot,\cdot|\bZ)$, one can resort to the usual nonparametric estimation of the conditional distribution, such as kernel or spline methods. 
However, it would suffer from the so-called curse of dimensionality, especially when the dimension of $\bZ$ is high. 
To avoid this issue, 
we develop a 
parametric regression framework for modeling the conditional quantile correlation coefficient. 
Specifically, 
we directly model qcc via a generalized linear  model, 
\begin{align}
	\label{eqn:mod}
	\phi_F\left(\operatorname{qcc}(\btau|\bZ)\right)
	=\bZ_3^\top \bgamma(\btau),  \quad \btau\in \mathbb{D}, 
\end{align}
where the link function  $\phi_F(r)=\frac{1}{2}\log\left(\frac{1+r}{1-r}\right)$ 
is the well-known Fisher z-transformation function for $r\in[-1,1]$, 
and $\mathbb{D}$ is a compact subset of the unit square $(0,1)^2$,  
$\bZ_3$  is a $(p_3+1)\times 1$ subvector of $\bZ$ that may overlap with $\bZ_1$ and $\bZ_2$, including 1 as the first element, 
and the $(p_3+1)\times 1$ regression coefficient vector $\bgamma(\btau)$ is an unknown function of $\btau$.

\noindent\textbf{Remark~2.} 
Since qcc lies between -1 and 1, it is natural and convenient to use Fisher z-transformation function. Certainly, one can use other transformations to link qcc to the regression part
in the right hand side of model~\eqref{eqn:mod}. 

\subsection{Unbiased estimating equation estimator}
\label{ss:EE}

Our main interest is to estimate and make statistical inference for the unknown $\bgamma(\btau)$ in model~\eqref{eqn:mod}.  
Our proposal for estimating $\bgamma(\btau)$ is motivated by the relationship between the conditional qcc and the direct model in ~\eqref{eqn:mod}. 
Specifically, 
by 
\eqref{eqn:cqcc-equiv} and \eqref{eqn:mod}, simple algebra can obtain the estimating equation 
for a pre-specified $\btau\in \mathbb{D}$, 
\begin{align}
	\label{eqn:ee-population}
	\operatorname{E}
	\left[
	A(\btau|\bZ) -g\left(\bZ_3^\top\bgamma, \btau\right)\bigg|\bZ
	\right] =0, 
\end{align} 
where 
$A(\btau|\bZ)=I\left\{Y_1\leq Q_{\tau_1}(Y_1|\bZ), Y_2\leq Q_{\tau_2}(Y_2|\bZ)\right\}$ is an indicator function whether the bivariate outcomes $(Y_1, Y_2)$ are located below the corresponding quantile levels simultaneously,  
$g(u,\btau)=g_0\left(\phi_F^{-1}(u),\tau\right)$
with 
$g_0(u, \btau)=\tau_1\tau_2+u\sqrt{(\tau_1-\tau_1^2)(\tau_2-\tau_2^2)}$,
and 
$
\phi_F^{-1}(u)=\frac{\exp(2u)-1}{\exp(2u)+1}
$ 
is the inverse function of the Fisher z-transformation function $\phi_F(\cdot)$.

For $n$ IID samples $\{(Y_{i1}, Y_{i2}, \bZ_{i}): i=1,\ldots,n\}$ from $(Y_1, Y_2, \bZ)$,   
we define  
$$
\widehat{A}(\btau|\bZ_i)=I\left\{Y_{i1}\leq \widehat{Q}_{\tau_1}(Y_{i1}|\bZ), Y_{i2}\leq \widehat{Q}_{\tau_2}(Y_{i2}|\bZ)\right\}, 
$$
where $\widehat{Q}_{\tau_j}(Y_{ij}|\bZ)=\bZ_j^\top\widehat{\bbeta}_j(\tau_j)$, 
$j=1,2$, are consistent estimators for the marginal quantile regressions in \eqref{eqn:mod-marginal}. 
By the consistency of $\widehat{\bbeta}_j$, 
we propose
a finite-sample analog of \eqref{eqn:ee-population}, 
\begin{align}
	\label{eqn:ee-sample}
	U_n(\widehat{\bbeta}, \bgamma, \btau)
	=
	\frac{1}{n}\sum_{i=1}^n \bZ_{3i} \left[\widehat{A}(\btau|\bZ_{i})
	-g(\bZ_{3i}^\top\bgamma, \btau)
	\right],	
	\quad 
	\btau\in \mathbb{D},  
\end{align}
which is asymptotically unbiased at $\bgamma_0(\btau)$ 
under model \eqref{eqn:mod}. 
Then, 
we propose an estimating equation estimator
$\widehat{\bgamma}(\btau)$
of  $\bgamma_0(\btau)$ by solving 
$
U_n(\widehat{\bbeta}, \bgamma, \btau)=0. 
$

\subsection{Asymptotic properties}
\label{ss:asymptotic}
 
Throughout this paper, to avoid ambiguity, 
we use subscript $i$ for $n$ samples $i=1,\ldots,n$, and subscript $j$ for two marginals $j=1,2$. 
We impose the regularity conditions 
to establish the asymptotic properties of $\widehat{\bgamma}(\btau)$. 
\begin{description}
	\item[(C1)] 
	(Data).  $\{(Y_{i1}, Y_{i2}, \bZ_{i}): i=1,\ldots,n\}$ are independent and identically distributed (IID) replicates from $(Y_1, Y_2, \bZ)$, 
	where the covariates $\bZ$ is uniformly bounded.
	
	\item [(C2)]
	(Differentiability and boundedness of marginal densities). 
	(i) The marginal densities $f_j(y_j|\bZ_j)$ for $j=1,2$ are continuously differentiable with bounded derivatives. 
	(ii)
	The matrices 
	$B_j(\bbeta_{j0}(\tau_j))    =\operatorname{E}\left[f_j(\bZ_j^\top\bbeta_{j0}(\tau_j)|\bZ_j)\bZ_j\bZ_j^\top\right]$ for $j=1,2$ are positive definite and  invertible for any 
	$\btau=(\tau_1,\tau_2)^\top\in \mathbb{D}$, 
	where $\mathbb{D}\subset (0,1)\times (0,1)$ is a compact subset. 
	(iii) 
	For $j=1,2$, the true coefficients
	$\bbeta_{j0}\in\mathcal{B}_j$, 
	where $\mathcal{B}_j$ is a compact subset of $\mathbb{R}^{p_j+1}$.

	\item[(C3)] 
	(Boundedness and identifiability).
	(i) $\sup\limits_{\btau\in\mathbb{D}}\|\bgamma_0(\btau)\|$ is bounded from above zero. 
	(ii)
	There exists a constant $c>0$ such that 
	$\inf\limits_{\btau\in\mathbb{D}} eigmin J(\bgamma_0(\btau); \btau)>c$, 
	where $J(\bgamma; \btau)=\mbox{E}[g'(\bZ^\top\bgamma; \btau)\bZ\bZ^\top]$. 
	(iii) $\bgamma_{0}\in\Upsilon$, 
	where $\Upsilon$ is a compact subset of $\mathbb{R}^{p+1}$. 
	
	\item[(C4)]
	(Differentiability of joint distribution). 
	(i) Both $\partial h_1(y_1,y_2|\bZ)/\partial y_1$ and   $\partial h_2(y_1,y_2|\bZ)/\partial y_2$ 
	are continuously differentiable with bounded derivatives, where $h_j(y_1,y_2|\bZ)=\partial F(y_1,y_2|\bZ)/\partial y_j$ for $j=1,2$, are the partial derivative of $F(y_1,y_2|\bZ)$.  
	(ii) The matrix  $J(\bgamma_0(\btau); \btau)$ is positive definite and  invertible.  
\end{description}

Condition (C1) is the IID assumption of the data. 
Condition (C2) is standard for the asymptotic results of the marginal regression coefficients estimator $\widehat{\bbeta}_j(\cdot)$ in ordinal quantile regression theory \citep{koenker2005}.  
Condition (C3) imposes the boundedness 
and the identifiability  of $\bgamma(\btau)$ for $\tau\in\mathbb{D}$, which guarantees the unique solution of the estimating equation in \eqref{eqn:ee-population}. 
Condition (C4) requires differentiability of joint distribution and invertibility  of $J(\bgamma_0(\btau); \btau)$, which is used to establish the weak convergence of  $\widehat{\bgamma}(\btau)$.

\begin{theorem}
	\label{thm:asn}
	Under regularity conditions (C1)--(C4), as $n\rightarrow\infty$, 
	we have
	\begin{description}
		\item[(1)] (Consistency).
		$\sup\limits_{\btau\in\mathbb{D}} \| \widehat{\bgamma}(\btau)-\bgamma_0(\btau)\|\mathop{\longrightarrow}\limits^P 0$. 
		
		\item[(2)] (Weak convergence). 
			$\sqrt{n}\left(\widehat{\bgamma}(\btau)-\bgamma_0(\btau)\right)
			= J(\bgamma_0, \btau)^{-1} n^{-1/2}\sum\limits_{i=1}^n \xi_i(\bbeta_0,\bgamma_0, \btau) +o_P(1)$,
		uniformly on $\btau\in \mathbb{D}$, 
		where 	for $j=1,2$, 
		$\pi_{ji}(\tau_j)
		=\bZ_{ji} \psi_{\tau_j}
		\left(Y_{ji}-\bZ_{ji}^\top\bbeta_{j0}(\tau_j)\right)$,
		$$
		\xi_i(\bbeta,\bgamma, \btau)
		=
		\bZ_{3i}\left\{A_i(\bbeta,\btau|\bZ_i)-g(\bZ_{3i}^\top \bgamma,\btau)\right\}
		+
		\sum\limits_{j=1}^2 P_j(\bbeta, \btau)
		B_j^{-1}\left(\bbeta_{j0}(\tau_j)\right)  \pi_{ji}(\tau_j), 
		$$
		and 
		$ P_j\left(\bbeta_{10}(\tau_1), \bbeta_{20}(\tau_2)\right)=\operatorname{E}\left[\bZ_3 \bZ_j^\top h_j(\bZ_1^\top\bbeta_{10}(\tau_1), \bZ_2^\top\bbeta_{20}(\tau_2)|\bZ))\right]$.  
		Furthermore, 	$\sqrt{n}\left(\widehat{\bgamma}(\btau)-\bgamma_0(\btau)\right)$  converges weakly to a  zero-mean Gaussian process $G(\btau)$,
		whose covariance function is given by 
		$
		\Sigma(\btau, \btau')=
		J(\bgamma_0, \btau)^{-1}
		V(\btau,\btau')
		J(\bgamma_0, \btau')^{-1},
		$
		where $V(\btau,\btau')=\operatorname{E}\left[
		\xi_i(\bbeta_0,\bgamma_0, \btau)\xi_i(\bbeta_0,\bgamma_0, \btau')^\top
		\right]$. 
	\end{description}
\end{theorem}

\noindent\textbf{Remark 2} (Basic inference). 
The limiting distribution of the sample 
$\widehat{\operatorname{qcc}}(\btau)$ in \eqref{eqn:qcc-sample}
can be easily derived by taking $\bZ=1$, i.e., only with intercept term.   
From Theorem~\ref{thm:asn}, 
it is easy to obtain the asymptotic normality of $\widehat{\bgamma}(\btau)$ for each $\btau\in\mathbb{D}$,  that is, 
$
	\sqrt{n}\left(
	\widehat{\bgamma}(\btau)-\bgamma_0(\btau)
	\right)
	\mathop{\longrightarrow}\limits^\mathcal{D} 
	N(0, \Sigma(\btau,\btau)), 
$
where $\Sigma(\btau,\btau)$ is defined in Theorem~\ref{thm:asn}. 
In addition, for any finite collection of quantile indices 
$\{\btau_k: k\in I\}$ with $I=\{1,\ldots,m\}$, 
\begin{equation*}
	\left\{
	\sqrt{n}\left(\widehat{\bgamma}(\btau_k)-\bgamma_0(\btau_k)\right)
	\right\}_{k\in I}
	\mathop{\longrightarrow}\limits^\mathcal{D} 
	N\left(0, \left\{
	J(\bgamma_0, \btau_k)^{-1}
	V(\btau_k,\btau_l)
	J(\bgamma_0, \btau_l)^{-1}
	\right\}_{k,l\in I}
	\right), 
\end{equation*}
which gives the joint limit distribution of the proposed estimator for several quantiles. 
This implies that the result in Theorem~\ref{thm:asn} is stronger assuring that the entire empirical quantile correlation regression process $\widehat{\bgamma}(\cdot)$ 
asymptotically behaves continuously, 
enabling the uniform approximation of $\widehat{\bgamma}(\cdot)$ by a finite collection of quantile correlation regression quantiles $\widehat{\bgamma}(\btau_k)$, $k\in I$ for a suitably fine grid of quantile indices $\{\btau_k: k\in I\}$.  

\noindent\textbf{Remark 3} (Standard errors).  
$\Sigma(\btau,\btau)$ in Theorem~\ref{thm:asn} can be consistently estimated by 
$$
\widehat{\Sigma}(\btau,\btau')
=\widehat{J}_n(\widehat{\bgamma},\btau)^{-1} \widehat{V}_n(\widehat{\bbeta},\widehat{\bgamma}, \btau,\btau') \widehat{J}_n(\widehat{\bgamma},\btau')^{-1},
$$
where $\widehat{J}_n(\bgamma,\btau) 
= n^{-1}\sum\limits_{i=1}^n 
\bZ_{3i}\bZ_{3i}^\top 
g'(\bZ_{3i}^\top \bgamma,\btau)$, 
and 
$\widehat{V}_n(\bbeta,\bgamma, \btau, \btau')
=n^{-1}\sum\limits_{i=1}^n \widehat{\xi}_i(\bbeta,\bgamma,\btau)
\widehat{\xi}_i(\bbeta,\bgamma,\btau')^\top$, 
\begin{align*}
	\widehat{\xi}_i(\bbeta,\bgamma,\btau)
	=\bZ_{3i}\left\{A_i(\bbeta,\btau|\bZ_i)-g(\bZ_{3i}^\top \bgamma,\btau)\right\}
	+\sum\limits_{j=1}^2 \widehat{P}_j(\bbeta, \btau) \widehat{B}_j^{-1}\left(\bbeta_{j0}(\tau_j)\right) \pi_{ji}(\tau_j), 
\end{align*}
$$\widehat{P}_j\left(\bbeta,\btau\right)
=\sum\limits_{i=1}^n \bZ_{3i} \bZ_{ji}^\top K_{h_j}\left(\frac{Y_{ji}-\bZ_{ji}^\top\bbeta_j}{h_j}\right) I\left(Y_{3-j, i}\leq \bZ_{3-j,i}^\top\bbeta_{3-j}\right),
$$
$K_h(u)=K(u/h)/h$ with a kernel function $K(\cdot)$ and a bandwidth $h>0$, 
and 
$
\widehat{B}_j(\bbeta_j)=\frac{1}{2nb_j}\sum\limits_{i=1}^n \bZ_i\bZ_i^\top I\left(|Y_{ji}-\bZ_{ji}^\top\bbeta_j|\leq b_j\right)
$ 
with a bandwidth $b_j>0$, for $j=1,2$.  
In practice, we use the Epanechnikov kernel function $K(u)=0.75 (1-u^2)I(|u|\leq 1)$, and the optimal bandwidth  
$h_j=2.34 n^{-1/5} \min(sd(Y_j), IQR(Y_j)/1.34)$ 
by the Silverman rule of thumb, 
where $sd(Y_j)$ and $IQR(Y_j)$ are respectively the standard error and interquartile range of $Y_j$,  
and 
$b_j$ is selected by the Hall-Sheather rule~\citep{hall1988on} for $j=1,2$.  

\section{General Inference}
\label{s:test}

\subsection{Inference hypotheses and procedures}
\label{ss:inference}

In this section, we propose the general hypothesis testing for $\bgamma(\btau)$. 
It is convenient to embed our hypotheses in the following null hypothesis, 
\begin{align}
	\label{eqn:null}
	H_0: \bm{R}\bgamma(\btau)=\bzeta(\btau) \quad \text{for all $\btau\in \mathbb{D}$},
\end{align}
against the alternative hypothesis 
\begin{align}
	\label{eqn:alternative}
	H_1: \bm{R}\bgamma(\btau)\neq \bzeta(\btau) \quad \text{for some $\btau\in \mathbb{D}$},
\end{align}
where $\bm{R}$ is a known $q\times (p_3+1)$ matrix of rank $q$, $q\leq dim(\bgamma)=p_3+1$, 
and $\bzeta(\btau)\in\mathbb{R}^q$. 
This null hypothesis might entail the assertion that some subset of the covariates exert no influence on the conditional qcc of the response over a specified range of qcc levels, $\btau\in\mathbb{D}$. 

We focus on the basic inference process 
$	\nu_n(\btau) = \sqrt{n}\left[\bm{R}\widehat{\bgamma}(\btau) -\widehat{\bzeta}(\btau) \right],
$
and test statistics of the form 
$T_n=T(\nu_n(\cdot))$ derived from it, 
where $T(\cdot)$ is any continuous functional, 
and $\widehat{\bzeta}(\btau)$ is a consistent estimator of $\bzeta(\btau)$.  
Two popular examples of such statistics are the 
Kolmogorov-Smirnov (KS) and Cramer-von Misses (CvM) statistics, 
\begin{equation}
	\label{eq:test}
	\begin{aligned}
		T_n^{KS} 
		=
		\sup\limits_{\btau\in\mathbb{D}}\| \nu_n(\btau)\|_{\widehat{\Lambda}(\btau)},
		\quad \quad 
		T_n^{CvM} 
		=
		\int_{\mathbb{D}}\| \nu_n(\btau)\|_{\widehat{\Lambda}(\btau)}^2d\btau,
	\end{aligned}
\end{equation}
respectively, 
where 
$\|\bzeta\|_\Lambda = \sqrt{\bzeta^\top\Lambda\bzeta}$ 
for any $\bzeta\in\mathbb{R}^q$ and positive definite matrix $\Lambda$,  
the symmetric $\widehat{\Lambda}(\btau)$ is a consistent estimator of $\Lambda(\tau)$ uniformly  in $\tau$, 
and $\Lambda(\tau)$ is a positive definite symmetric matrix uniformly in $\tau$, e.g., $\Lambda(\tau)=I_q$ and $\Lambda(\tau)
=
\left[\operatorname{Var}(\nu_n(\tau))\right]^{-1}
=
\left[
\bm{R}\Sigma(\btau,\btau)\bm{R}^\top
\right]^{-1}$. 
Intuitively, one can reject the null hypothesis if the test statistic $T_n^{KS}$ or $T_n^{CvM}$ is large. 
The null hypothesis can be rejected if 
$
T_n>c_{\alpha},
$
where $c_{\alpha}$ is the critical value of $T_n$ at significant level $\alpha$. 
The formal inference results are presented in Section~\ref{ss:formal}, 
and the  critical values are obtained using the resampling procedure in Section~\ref{ss:resampling}. 

The following examples are some hypotheses that may be considered in this framework. 

\medskip
\noindent\textbf{Example~1} (Hypothesis of no effect). 
A basic hypothesis is the overall significance of  $\gamma^{(k)} (\btau)$ for $k=0,1,\ldots, p_3$. Specifically, 
the null hypothesis is
$H_0^{(k)}: 
\gamma^{(k)} (\btau)=0$ 
for all $\btau\in \mathbb{D}$. 
Note that 
$H_0^{(k)}$ corresponds to a test of independence for the baseline group when $k=0$, and tests whether 
the covariate effects have no impact on the dependence when $k=1,\ldots,p_3$. 
In this case, $\bm{R}=[0,\ldots,1,\ldots, 0]$ and $\bm{\zeta}(\cdot)=0$. 

The next example considers a hypothesis of constant covarite effects.  The alternative is that the covariate effect on dependence  varies across quantiles, which is of fundamental importance because it motivates structural models developed to account for varying effects. 

\medskip
\noindent\textbf{Example~2} (Constant effect hypothesis). 
The hypothesis of a constant effect on dependence is whether the covariate effects on dependence remain constant across quantiles. 
Then, 
$H_0^{(k)}: 
\gamma^{(k)} (\btau)=c_k$ 
for all $\btau\in \mathbb{D}$, 
where $c_k$ is an unknown constant. 
Rejection of $H_0^{(k)}$ suggests varying covariate effects at different levels of the outcome variables. 
This asserts that $\gamma^{(k)}$ is constant across all $\btau\in \mathbb{D}$. 
In this case, $\bm{R}=[0,\ldots,1,\ldots, 0]$ and $\bm{\zeta}(\cdot)=c_k$.

\subsection{Formal inference results} 
\label{ss:formal}

To establish the formal inference results,  
we maintain the following condition for inference.  
\begin{description}
	\item[(C5)] 
	(Condition for inference). 
	(i) $\bm{R}\bgamma(\btau)=\bzeta(\btau)$, 
	where $\bm{R}, \bzeta(\btau)$ are continuous and either (a) $\bzeta(\btau)=0$ for all $\btau$ or
	(b) $\bzeta(\btau)\neq 0$ for some $\btau$.  
	(ii) 
	$\sqrt{n}(\widehat{\bzeta}(\btau)-\bzeta(\btau))=n^{-1/2}\sum\limits_{i=1}^nd_i(\btau)+o_P(\btau)$ uniformly in $\mathbb{D}$, 
	where $d_i(\btau)$ is IID mean zero for each $\btau$, 
	and hence  either $\sqrt{n}(\widehat{\bzeta}(\btau)-\bzeta(\btau))$ converges weakly to (a) $W(\btau)\equiv 0$ if $\bzeta(\btau)=0$ for all $\btau$ or 
	(b) a zero-mean Gaussian process $W(\btau)$ for $\btau\in \mathbb{D}$ if $\bzeta(\btau)\neq 0$ for some $\btau$, where $W(\btau)$ has the covariance function $\Omega(\btau,\btau')=\operatorname{E}\left[
	d_1(\btau)d_1(\btau')\right]$. 
\end{description} 

The following results establish the limiting distribution of the test statistic $T_n$ under both the null  and alternative hypotheses.  
\begin{theorem}
	\label{thm:inference}
	Suppose that Conditions (C1)--(C5) hold.  
	Then, for any continuous functional $T(\cdot)$, 
	the following statements hold: 
	\begin{description}
		\item[(i)] (Limiting null distribution).
		Under 
		$H_0$ in \eqref{eqn:null},  
		$T_n=T(\nu_n(\cdot))$  converges weakly to the process $T_\infty = T\left(\bm{R} G(\btau)- W(\btau)\right)$,  
		where $G(\btau)$ is  a zero-mean  Gaussian process on $\mathbb{D}$ defined in Theorem~\ref{thm:asn}, and $W(\btau)$ is defined in condition (C5).  
		
		\item [(ii)] (Consistency). 
		Under 
		$H_1$ in \eqref{eqn:alternative},  $P(T_n>a)\rightarrow 1$ for any constant $a>0$.         
	\end{description}
	
\end{theorem}


To investigate the local power of the test, 
we derive the limiting distribution of the test statistic $T_n$ under a sequence of local alternatives that shrink toward to the null at rate $n^{-1/2}$. 
Specifically,
we consider the following local alternative hypothesis,
\begin{align}
	\label{eqn:H1a}
	H_{1n}:  \bm{R}\bgamma(\btau)-\bzeta(\tau)=n^{-1/2}\bm{\varsigma}(\btau),  
	\quad \text{for some $\btau \in \mathbb{D}$, } 
\end{align}
where $\bm{\varsigma}(\tau)$ is $q$-vector of continuous and bounded function on $\mathbb{D}$. 

\begin{theorem}
	\label{thm:local-alter}
	(Local power). 
	Assume that Conditions (C1)--(C5) hold.  
	Then, 
	under the local alternative hypothesis $H_{1n}$ in \eqref{eqn:H1a},  
	for any continuous functional $T(\cdot)$, 
	$T_n=T\left\{\nu_n(\cdot)\right\}$ converges weakly to the process 
	$
	T\left\{\bm{R}G(\btau)-W(\btau) +\bm{q}(\btau)\right\}
	$, 
	where  $G(\btau)$ and $W(\btau)$  are respectively defined in Theorem~\ref{thm:asn} and condition (C5),  
	and $\bm{q}(\btau)=\bm{R}\left(\bm{R}^\top\bm{R}\right)^{-1}
	\bm{R}^\top\bm{\varsigma}(\btau)$. 
\end{theorem}

Theorem~\ref{thm:local-alter} shows 
that under $n^{-1/2}$ local departures, 
the limit distribution of $T_n$ is the functional of a shifted Gaussian process
$\bm{R}G(\btau)-W(\btau) +\bm{q}(\btau)$, 
and thus  $T_n$  retains nontrivial power against local misspecification. 
Besides,  
the rejection probability increases monotonically with the magnitude of $\bm{q}(\btau)$, 
providing higher power for larger deviations.

Next, we briefly go through our testing Examples 1-2 and state the scores for each of them. 
\medskip
\noindent\textbf{Example~1} (Continued. Test of no effect). 
To test the null hypothesis $H_{01}^{(k)}: \gamma^{(k)} (\btau)=0$ for all $\btau\in \mathbb{D}$,  
it is natural to use the average of $\widehat{\bgamma}(\btau)$ across $\btau\in\mathbb{D}$, i.e., 
$
\widehat{\bm\varrho}=\frac{1}{|\mathbb{D}|}\int_{\mathbb{D}} \widehat{\bgamma}(\btau)d \btau,
$
where $|\mathbb{D}|$ is the volume of $\mathbb{D}$. 
By Theorem~\ref{thm:asn} and the continuous mapping theorem, it is easy to show that $\sqrt{n}(\widehat{\bm\varrho}-\bm\varrho_0)\mathop{\longrightarrow}\limits^\mathcal{D} N(0, \bm{\Sigma}_\varrho)$, 
where 
$
\bm\varrho_0=\frac{1}{|\mathbb{D}|}\int_{\mathbb{D}} \bgamma_0(\btau)d \btau,
$
and 
$
\bm{\Sigma}_\varrho=\operatorname{E}\left[
\frac{1}{|\mathbb{D}|}\int_{\mathbb{D}} 
J(\bgamma_0, \btau)^{-1}
\xi_i(\bbeta_0,\bgamma_0,\btau)d \btau\right]^{\bigotimes 2}. 
$
Then, a Wald-type statistic can be constructed for testing $H_{01}^{(k)}$ by 
$ \mathcal{T}_{1,W}^{(k)}=\sqrt{n}\widehat{\bm\varrho}^{(k)}/\sqrt{\widehat{\bm\Sigma}_{\varrho}^{(k,k)}},
$
whose limiting distribution is the standard normal distribution $N(0,1)$, 
where $\widehat{\varrho}^{(k)}$ is the $k$th element of $\widehat{\bm\varrho}$, 
and $\widehat{\bm\Sigma}_{\varrho}^{(k,k)}$ is the corresponding diagonal element of 
$
\widehat{\bm\Sigma}_{\varrho}
=n^{-1}\sum\limits_{i=1}^n\left[
\frac{1}{|\mathbb{D}|}\int_{\mathbb{D}} 
\widehat{J}(\widehat{\bgamma}, \btau)^{-1}\widehat{\xi}_i(\widehat{\bbeta},\widehat{\bgamma},\btau)d \btau\right]^{\bigotimes 2},
$
which is a consistent estimator of $\bm\Sigma_{\varrho}$. 

Basically, $\mathcal{T}_{1,W}^{(k)}$ is the averaged association for $\btau\in\mathbb{D}$, hence it is suitable for testing the overall strength of association. 
However, $\mathcal{T}_{1,W}^{(k)}$ may lack sensitivity when covariate effect only influences the association for a small range of $\btau$ or when the effect changes sign for different $\btau$. 
Therefore, we propose two typical types of test statistics for testing $H_{01}^{(k)}$, 
\begin{align*}
	\mathcal{T}_{1,KS}^{(k)}
	&=
	\sup\limits_{\btau\in\mathbb{D}}\| \bm{R}_k\widehat{\bgamma}(\btau)\|_{\widehat{\Lambda}(\btau)}
	=
	\sup\limits_{\btau\in\mathbb{D}} 
	\left|
	\frac{\sqrt{n} \widehat{\bgamma}^{(k)}(\btau)}{\sqrt{\widehat{\bm\Sigma}^{(k,k)}(\btau,\btau)}}
	\right|,\\
	\mathcal{T}_{1,CvM}^{(k)}
	&=
	\int_{\mathbb{D}}\|
	\bm{R}_k\widehat{\bgamma}(\btau)\|_{\widehat{\Lambda}(\btau)}^2d\btau
	=
	\int_{\mathbb{D}}
	\left[
	\frac{\sqrt{n} \widehat{\bgamma}^{(k)}(\btau)}{\sqrt{\widehat{\bm\Sigma}^{(k,k)}(\btau,\btau)}}
	\right]^2 d\btau,
\end{align*}
where 
$\widehat{\Lambda}(\btau)=\left[
\bm{R}_k\widehat{\Sigma}(\btau,\btau)\bm{R}_k
\right]^{-1}$,
and 
$\bm{R}_k=[0,\ldots,1,\ldots, 0]$. 
Both 
$\mathcal{T}_{1,KS}^{(k)}$  and $\mathcal{T}_{1,CvM}^{(k)}$ are omnibus for all types of departures from the null hypothesis and could by more powerful than $\mathcal{T}_{1,W}^{(k)}$ when covariate effects vary with $\btau$. 

\medskip
\noindent\textbf{Example~2}  (Continued. Test of constant effect). 
For an unknown constant $c_k$, 
to test the null hypothesis 
$H_{02}^{(k)}: \gamma^{(k)} (\btau)=c_k$ for all $\btau\in \mathbb{D}$,  
we propose two test statistics, 
\begin{align*}
	\mathcal{T}_{2,KS}^{(k)}
	&=
	\sup\limits_{\btau\in\mathbb{D}}\| \bm{R}_k\left(\widehat{\bgamma}(\btau)-\widehat{\bm\varrho}\right)\|_{\widehat{\Lambda}(\btau)}
	=
	\sup\limits_{\btau\in\mathbb{D}} 
	\left|
	\frac{\sqrt{n}\left(\widehat{\bgamma}^{(k)}(\btau)-\widehat{\bm\varrho}^{(k)}\right)}{\sqrt{\widehat{\Pi}^{(k,k)}(\btau,\btau)}}
	\right|,\\
	\mathcal{T}_{2,CvM}^{(k)}
	&=
	\int_{\mathbb{D}}
	\| \bm{R}_k\left(\widehat{\bgamma}(\btau)-\widehat{\bm\varrho}\right)\|_{\widehat{\Lambda}(\btau)}
	d\btau
	=
	\int_{\mathbb{D}}
	\left|
	\frac{\sqrt{n}\left(\widehat{\bgamma}^{(k)}(\btau)-\widehat{\bm\varrho}^{(k)}\right)}{\sqrt{\widehat{\Pi}^{(k,k)}(\btau,\btau)}}
	\right|^2
	d\btau,
\end{align*}
where 
$\widehat{\Lambda}(\btau)=\left[
\bm{R}_k\widehat{\Pi}(\btau,\btau)\bm{R}_k
\right]^{-1}$, 
$\bm{R}_k=[0,\ldots,1,\ldots, 0]$, 
and $\widehat{\Pi}(\btau,\btau) $ is a consistent estimator for the asymptotic covariance matrix of $\sqrt{n}(\widehat{\bgamma}(\btau)-\widehat{\bm\varrho})$. 

\subsection{Critical values by multiplier bootstrap}
\label{ss:resampling}

As showed in the above sections, the asymptotic distribution of  $T_n=T(v_n(\cdot))$ relies on the underlying data generating process and the continuous functional $T(\cdot)$ itself. 
Moreover, the complicated covariance structure of $T_\infty$ in Theorem~\ref{thm:inference} makes critical values not readily available. 
To this end, we propose to calculate critical values by the Gaussian multiplier bootstrap, which resamples the linear approximations of the empirical inference processes. 

We propose to approximate the asymptotic representation of $\sqrt{n}\left(\widehat{\bgamma}(\btau)-\bgamma_0(\btau)\right)$ by that of 
\begin{equation}
	\label{eq:G-boot}
	\widehat{G}_n^*(\btau)
	=n^{-1/2} \widehat{J}_n(\widehat{\bgamma},\btau)^{-1}
	\sum\limits_{i=1}^n  \widehat{\xi}_i(\widehat{\bbeta},\widehat{\bgamma},\btau)V_i,
\end{equation}
where $\{V_i, i=1,\ldots,n\}$ is a sequence of IID random variables with zero mean, unit variance (e.g., $V_i\sim N(0,1)$), 
and independent of the original data $\{(Y_{i1}, Y_{i2}, \bZ_i): i=1,\ldots,n\}$. 
Similarly, $W(\btau)$ for $\btau\in \mathbb{D}$ if $\bzeta(\btau)\neq 0$ for some $\btau$ can be approximated by 
$\widehat{W}_n^*(\btau)
=n^{-1/2}\sum\limits_{i=1}^n  \widehat{d}_i(\btau)V_i$. 
Then, the bootstrapped version of the proposed test statistic $T_n=T(\nu_n(\cdot))$ is simply given by  
$T_n^*=T\left\{\bm{R}\widehat{G}_n^*(\btau)-\widehat{W}_n^*(\btau)\right\}$. 
For example, the bootstrapped versions of KS and CvM type statistics are respectively given by 
		$T_n^{KS,*} 
		=
		\sup\limits_{\btau\in\mathbb{D}}
		\| \bm{R}\widehat{G}_n^*(\btau)\|_{\widehat{\Lambda}(\btau)}
		$,
		and
		$
		T_n^{CvM,*} 
		=
		\int_{\mathbb{D}}\| \bm{R}\widehat{G}_n^*(\btau)\|_{\widehat{\Lambda}(\btau)}^2d\btau.
		$
By replicating $B$ times, 
one can obtain $\{T_n^{*b}: b=1,\ldots, B\}$, 
and its empirical distribution estimator 
$\widehat{T}_n^{*b}(t)=B^{-1}\sum\limits_{b=1}^B I(T_n^{*b}\leq t)$. 
Then, the asymptotic critical value is obtained as the $(1-\alpha)$-th quantile of 
$\widehat{T}_n^{*b}(t)$, i.e.,  
\begin{equation}
	\label{eq:critical-boot}
	c_{n,\alpha}=\inf\{t: \widehat{T}_n^{*b}(t)\geq 1-\alpha \}. 
\end{equation}
The significance level $\alpha$ test rejects the null hypothesis if $T_n>c_{n,\alpha}$. 
Algorithm~1 summarizes the multiplier bootstrap procedure to  facilitate this feasible and practical implementation. 

\fbox{
	\begin{minipage}{\dimexpr\textwidth-8\fboxsep-1\fboxrule\relax} 
		\parbox{1\textwidth}{%
			\textbf{Algorithm~1:} 
			Simulated Test Procedure of $T_n$.  
			\begin{center} 
				\begin{description}
					\item[1]  
					Generate IID random variables 
					$\{V_i^b: i=1,..,n\}$ from $N(0,1)$. 
					
					\item[2]  
					Calculate $T_n^{*b}=T\left(
					\bm{R}\widehat{G}_n^{*b}(\btau)-\widehat{W}_n^{*b}
					\right)$, 
					where 		
					$
					\widehat{W}_n^*(\btau)
					=n^{-1/2}\sum\limits_{i=1}^n  \widehat{d}_i(\btau)V_i^b,
					$		   
					and 
					$
					\widehat{G}_n^{*b}(\btau)
					=n^{-1/2} \widehat{J}_n(\widehat{\bgamma},\btau)^{-1}
					\sum\limits_{i=1}^n  \widehat{\xi}_i(\widehat{\bbeta},\widehat{\bgamma},\btau)V_i^b$.

					\item[3]
					Repeat Steps~1--2 with $B$ times (e.g., B=1000) to obtain  $T_n^{*1}, \cdots, T_n^{*B}$. 
					The critical value $c_{n,\alpha}$ 
					is obtained by \eqref{eq:critical-boot}, 
					and the p-value is  
					$\widehat{p}_n =B^{-1}\mathop{\sum}\limits_{b=1}^{B}
					I\{T_n^{*b}\geq T_n\}$.   	
				\end{description}
			\end{center}
		}%
	\end{minipage}
}%

The intuition behind this simulated procedure is the fact that when the observations are independent, 
the asymptotic linearization of $\sqrt{n}\left(\widehat{\bgamma}(\btau)-\bgamma_0(\btau)\right)$ in Theorem~\ref{thm:asn} can be approximated by its perturbed version.  
The following theorem presents the formal justification of the multiplier bootstrap procedure. 
\begin{theorem}
	\label{thm:boot}
	Under conditions in Theorem~\ref{thm:inference}, 
	as $n\longrightarrow \infty$, 
	$\widehat{G}_n^*(\btau)$ and $\widehat{W}_n^*(\btau)$ converge weakly under the bootstrap law (i.e., conditional on the original data) to the process $G(\btau)$ and $W(\btau)$, respectively, 
	where  $G(\btau)$ and $W(\btau)$ are defined in Theorem~\ref{thm:asn}. 
	Additionally, for any continuous functional $T(\cdot)$, 
	we have that $T_n^*=T\left\{\bm{R}\widehat{G}_n^*(\btau)-\widehat{W}_n^*(\btau)\right\}$
	converges weakly to $T\left\{\bm{R} G(\btau)-W(\btau)\right\}$ under the bootstrap law. 
\end{theorem}

\section{Monte Carlo Simulations}
\label{s:sim} 

We evaluate the performance of the proposed method using a series of simulations, 
including the accuracy of estimation, the Type I error, and the power for detecting differences.

\subsection{Accuracy of estimation}
\label{ss:estimation}

We consider three different scenarios with different dependence structures of $(u_1, u_2)$:   
\begin{description}
	\item[(S1)]
	$(u_1, u_2)|\bZ\sim C_{\operatorname{FGM}}\left(u_1,  u_2; \phi_F^{-1}(\theta_{FGM,0} + \theta_{FGM,1} Z_2) \right)$, 
	
	\item[(S2)] 
	$(u_1, u_2)|\bZ\sim C_\Pi(u_1,  u_2) I(Z_2=0)
	+C_{\operatorname{Gumbel}}(u_1, u_2; \theta_G) I(Z_2=1)$, 
	
	\item[(S3)]
	$(u_1, u_2)|\bZ\sim C_{\operatorname{Clayton}}(u_1,  u_2; \theta_C) I(Z_2=0)+C_{\operatorname{Gumbel}}(u_1, u_2; \theta_G) I(Z_2=1)$,
	
\end{description}
where 
Clayton, Gumbel  and Farlie-Gumbel-Morgenstern (FGM) copulas are respectively 
\begin{align*}
	C_{\operatorname{Clayton}}(u_1, u_2; \theta_C)
	&=
	(u_1^{-\theta_C}+u_2^{-\theta_C}-1)^{-1/\theta_C},\\
	C_{\operatorname{Gumbel}}(u_1, u_2; \theta_G)
	&=
	\exp\left(-\left[(-\ln u_1)^{-\theta_G} + (-\ln u_2)^{-\theta_G}\right]^{-1/\theta_G}\right), \\
	C_{\operatorname{FGM}}(u_1, u_2; \theta_{FGM})
	&=
	u_1u_2+\theta_{FGM}\sqrt{u_1u_2(1-u_1)(1-u_2)},
\end{align*}
and $C_\Pi(u_1, u_2)$ is an independent copula. 
We set $\theta_C=0.857$, $\theta_G=1.429$ and $\theta_{FGM,0}=0.3$ for corresponding to a Kendall's $\tau$ coefficient of 0.3, 
and $\theta_{FGM,1}=0.15$. 
Then we generate bivariate outcomes $(y_1,y_2)$ by 
\begin{align*}
	\begin{cases}
		y_1
		&=\Phi^{-1}(u_1, 0, 0.2^2) +0.5 Z_1 + \left(0.5+\Phi^{-1}(u_1, 0, 0.2^2)\right)Z_2,  \\
		y_2
		&=\Phi^{-1}(u_2, 0, 0.3^2) -0.5 Z_1 + 0.5Z_2,  
	\end{cases}
\end{align*}
where two covariates $Z_1$ and $Z_2$ are independently generated from a standard normal distribution truncated at -2 and 2,  
and a Bernoulli distribution with successful probability 0.5,  respectively. 
Here, 
$\Phi^{-1}(u, \mu, \sigma^2)$ denotes the inverse of a normal distribution $\Phi(u, \mu, \sigma^2)$ with mean $\mu$ and standard deviation $\sigma^2$. 
Then, this DGP provides the qcc regression model,
$
\phi_F\left(\operatorname{qcc}(\btau|\bZ)\right)
=\gamma_0(\btau) + \gamma_1(\btau) Z_1 + \gamma_2(\btau) Z_2,
$
for any $\btau=(\tau_1, \tau_2)\in\mathbb{D}$, 
where the true parameter  
is respectively 
\begin{description}
	\item[(S1)] $\bgamma_0(\btau)=[0.3, 0, 0.15]^\top$, 
	
	\item[(S2)]  $\bgamma_0(\btau)=[0, 0, \varpi(\btau, C_{\operatorname{Gumbel}})]^\top$, 
	
	\item[(S3)] 
	$\bgamma_0(\btau)=[\varpi(\btau, C_{\operatorname{Clayton}}), 0, \varpi(\btau, C_{\operatorname{Gumbel}})-\varpi(\btau, C_{\operatorname{Clayton}})]^\top$, 
\end{description}
where 
$
\varpi(\btau, C)=\phi_F\left(\frac{C(\tau_1, \tau_2; \theta)-\tau_1\tau_2}{\sqrt{(\tau_1-\tau_1^2)(\tau_2-\tau_2^2)}}\right)
$
for any copula function $C(\tau_1,\tau_2;\theta)$ with parameter $\theta$. 
In these setups, 
scenario S1 is the situation where a covariate impacts the overall strength of association but not the association structure, 
scenarios S2  and S3 influence both the patterns  of the association strength and structure. 

\begin{figure}
	\centering
	\begin{tabular}{ccc}
		\makebox[0pt][r]{\makebox[20pt]{\raisebox{40pt}
				{\rotatebox[origin=l]{90}{S1}}}}
		\includegraphics[height=0.25\textheight,width=0.3\textwidth]
		{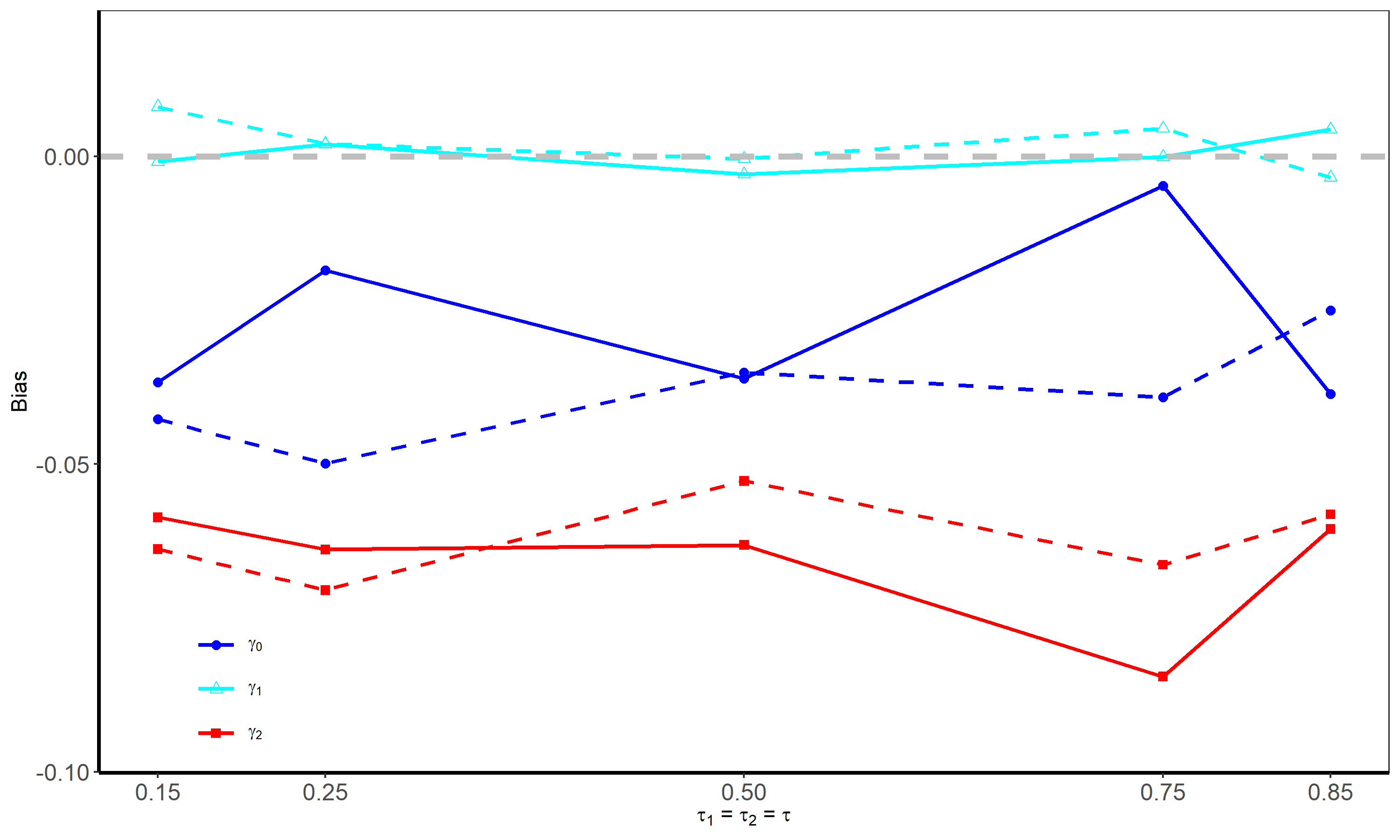} &
		\includegraphics[height=0.25\textheight,width=0.3\textwidth]
		{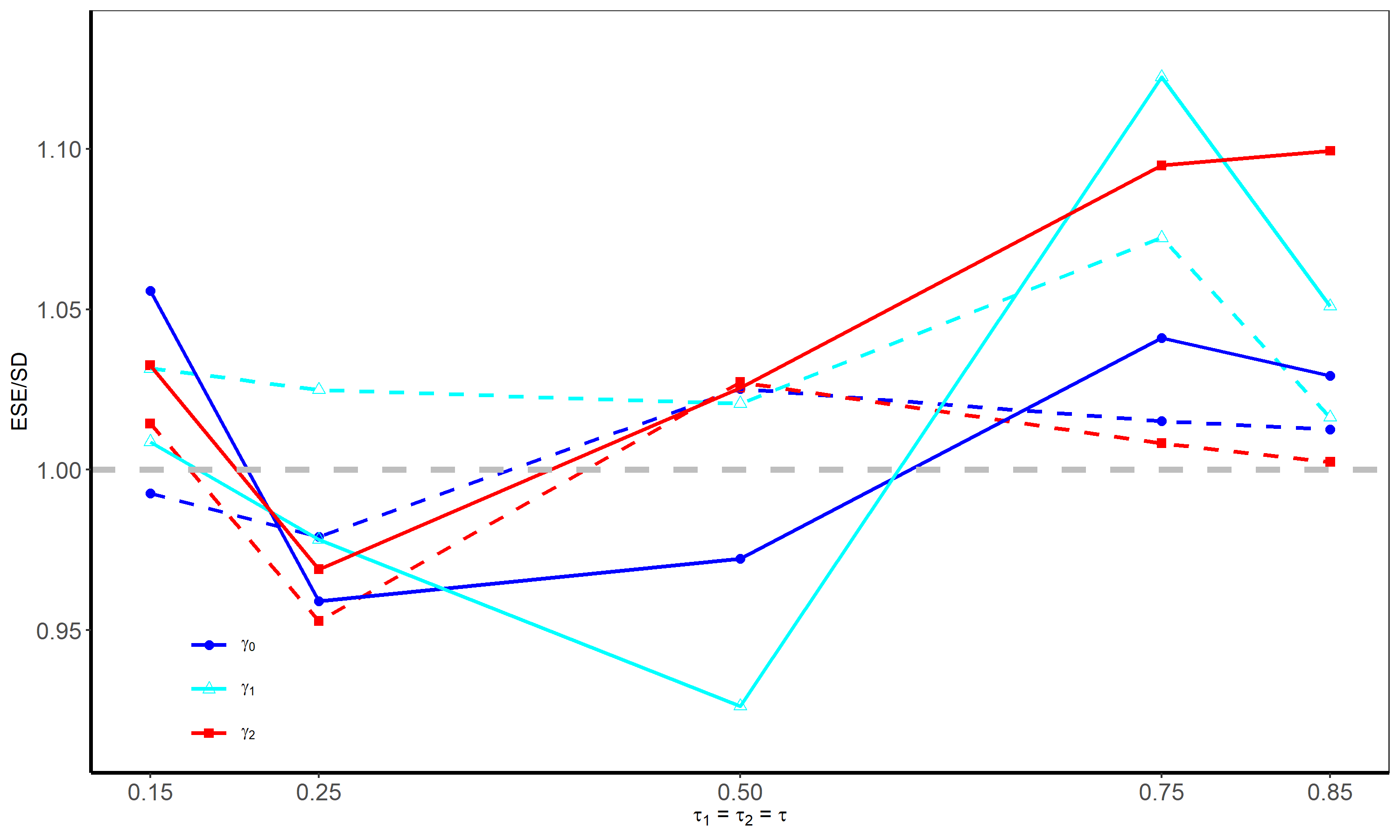} &
		\includegraphics[height=0.25\textheight,width=0.3\textwidth]
		{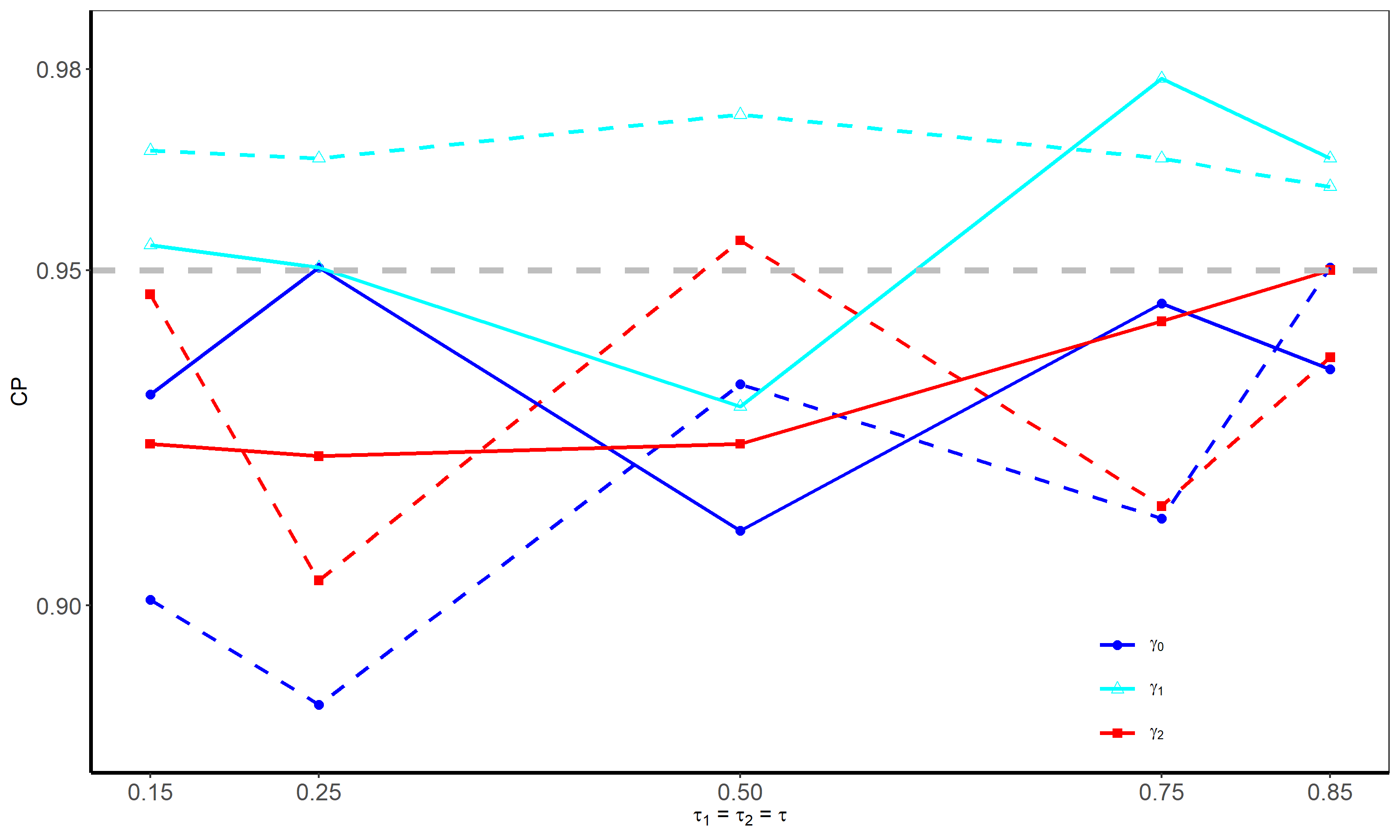}\\
		\makebox[0pt][r]{\makebox[20pt]{\raisebox{40pt}
				{\rotatebox[origin=l]{90}{S2}}}}
		\includegraphics[height=0.25\textheight,width=0.3\textwidth]
		{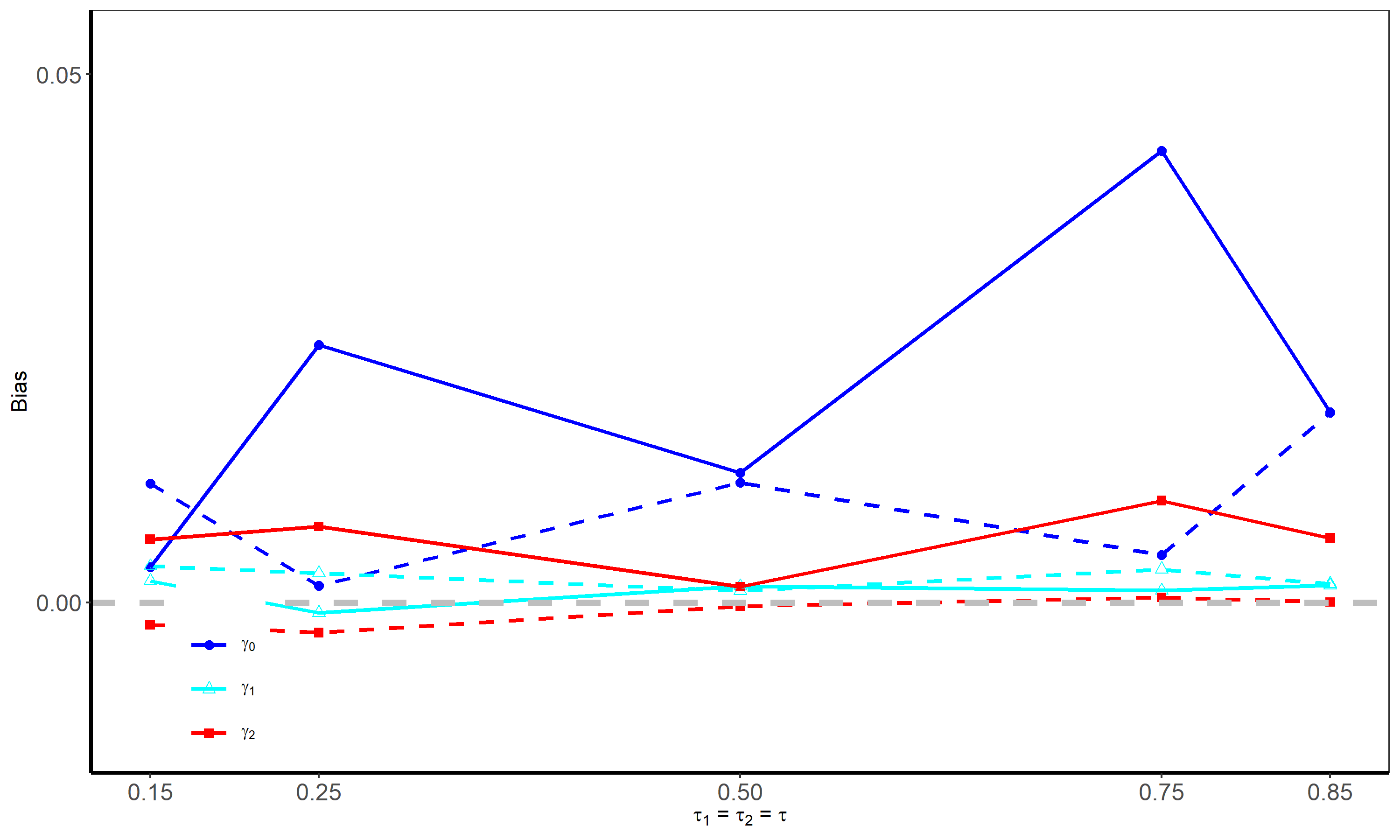} &
		\includegraphics[height=0.25\textheight,width=0.3\textwidth]
		{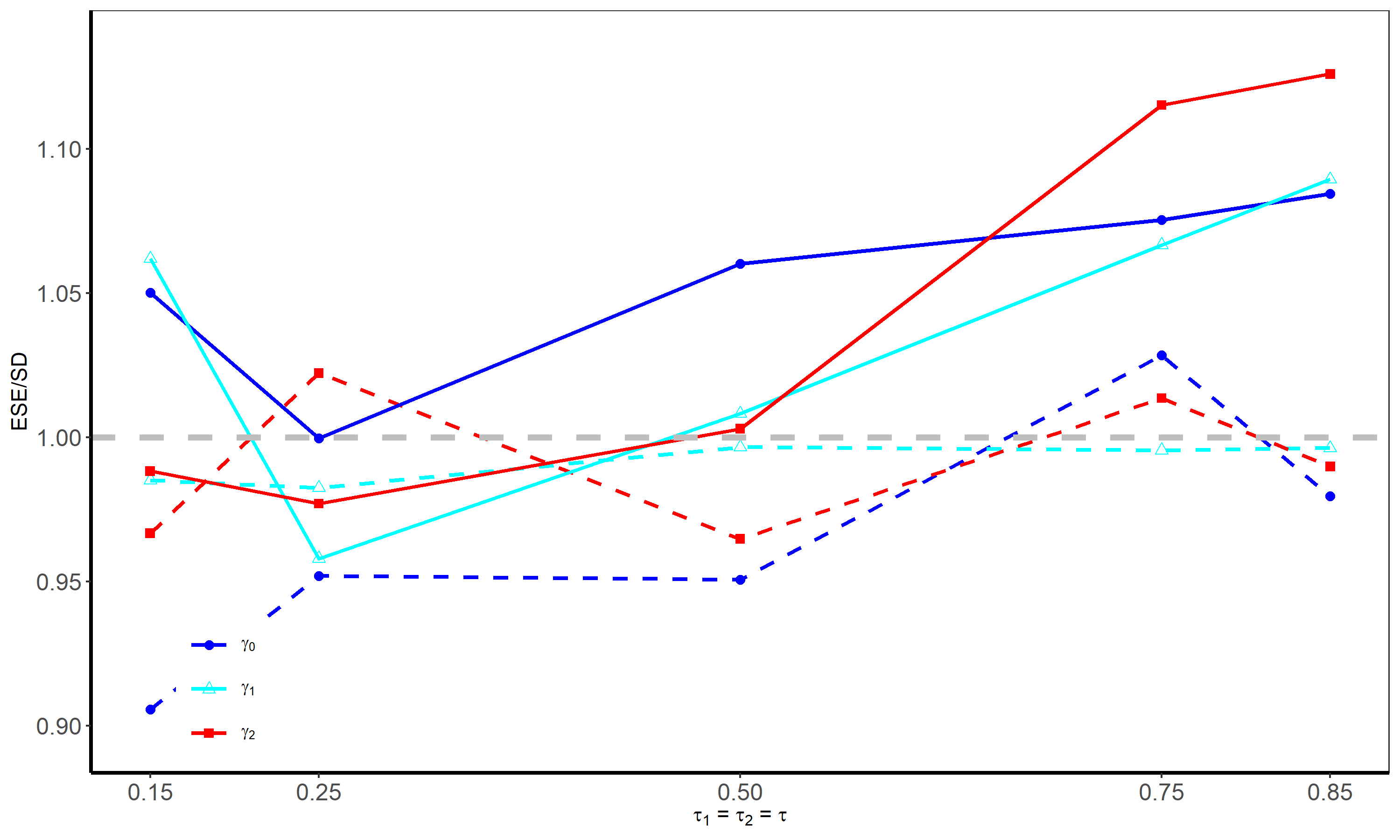} &
		\includegraphics[height=0.25\textheight,width=0.3\textwidth]
		{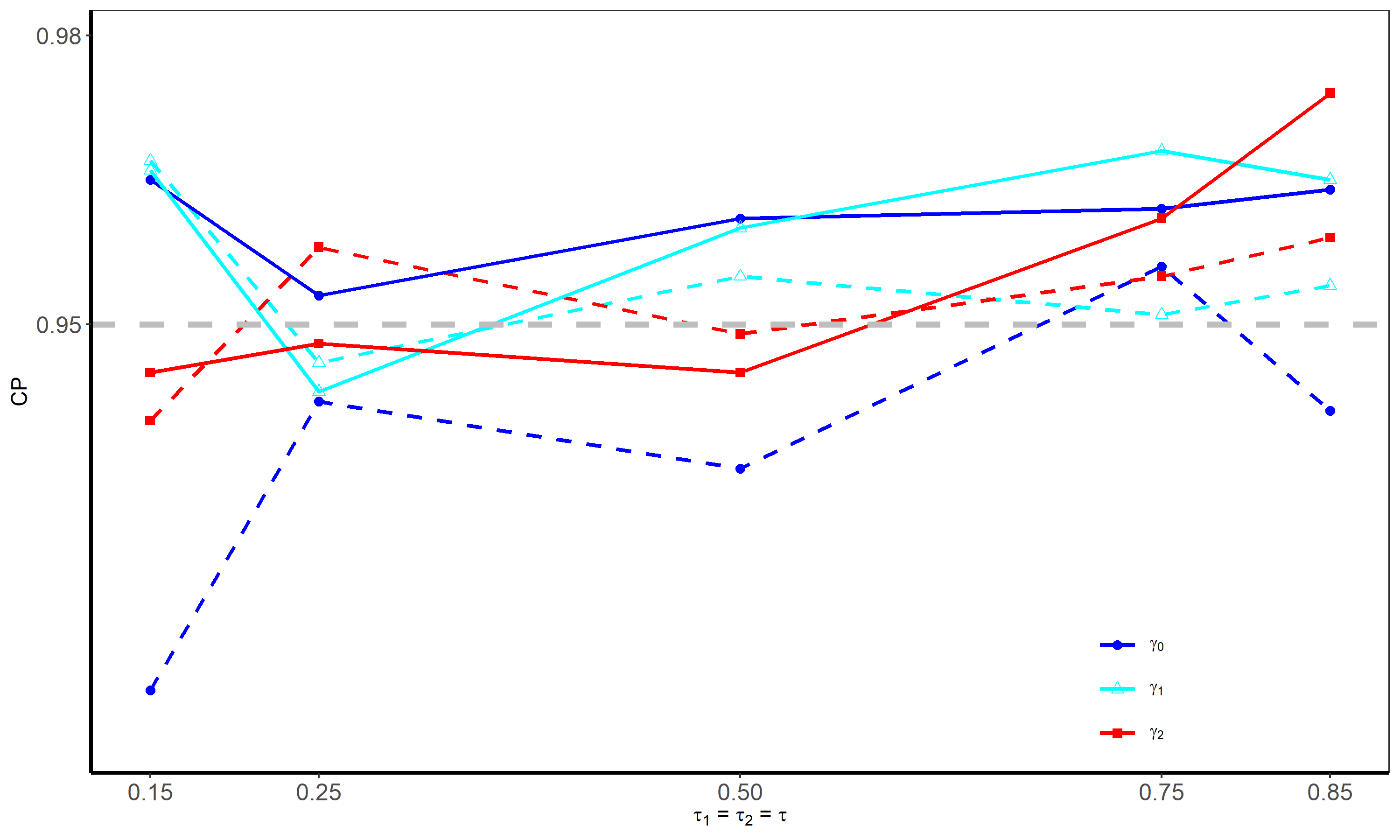}\\
		\makebox[0pt][r]{\makebox[20pt]{\raisebox{40pt}
				{\rotatebox[origin=l]{90}{S3}}}}
		\includegraphics[height=0.25\textheight,width=0.3\textwidth]
		{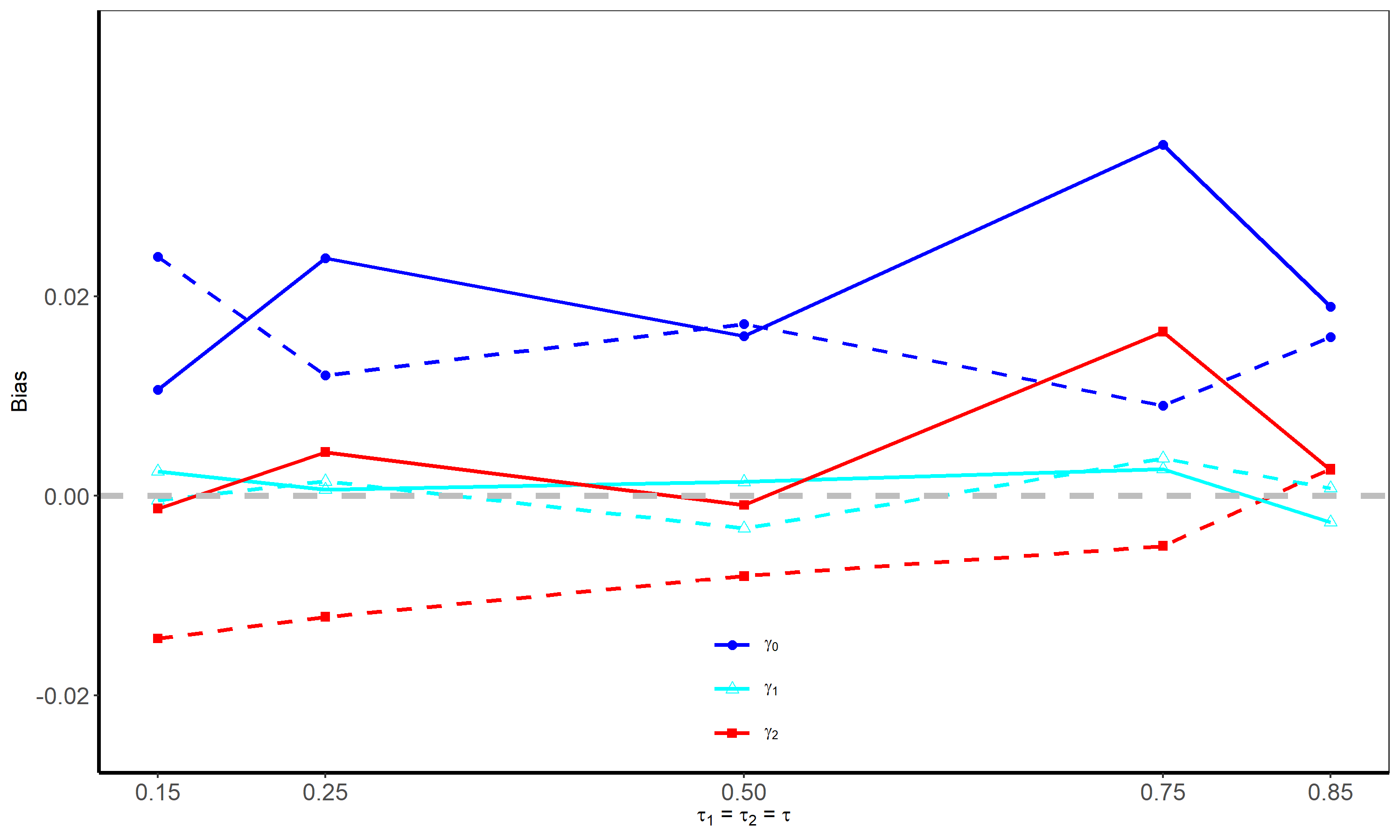} &
		\includegraphics[height=0.25\textheight,width=0.3\textwidth]
		{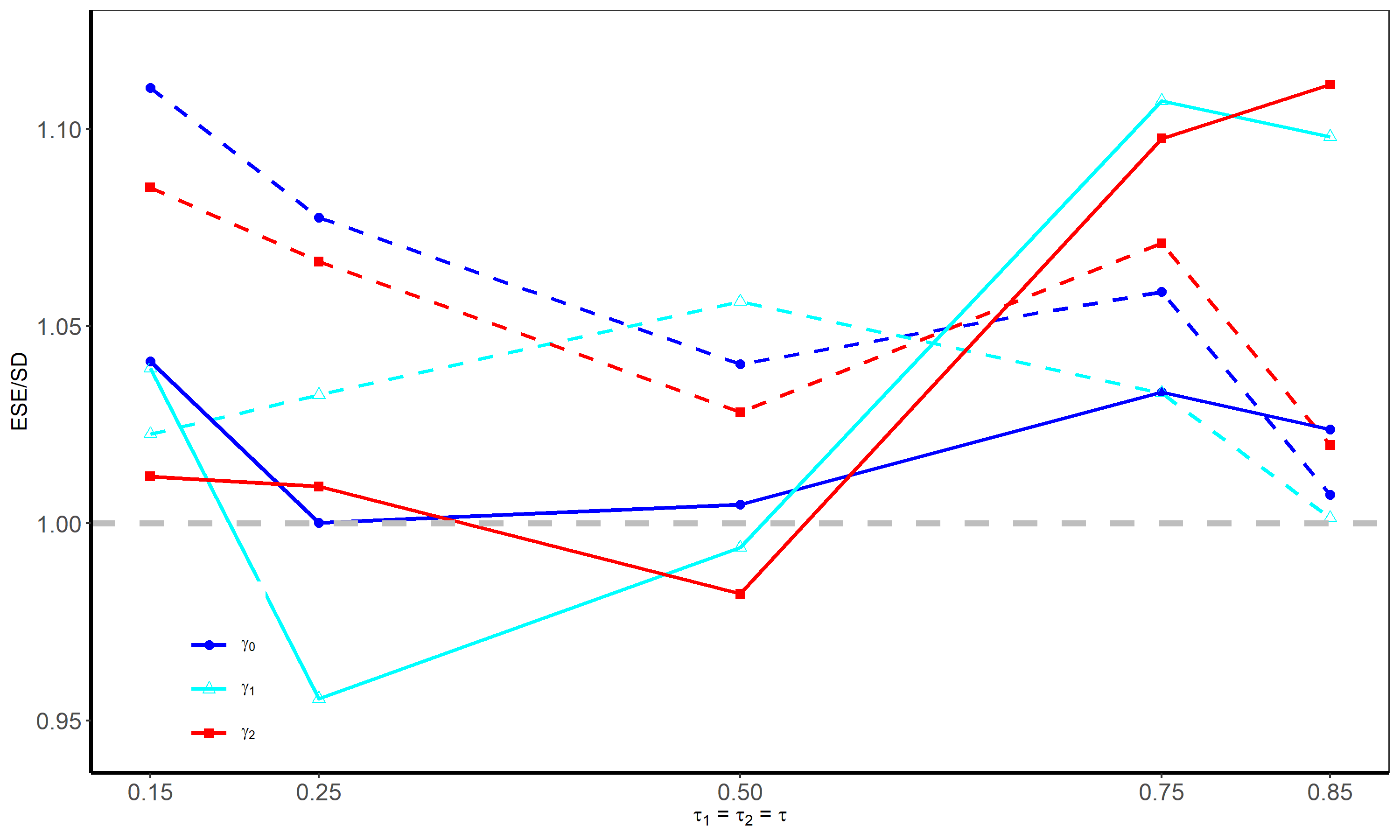} &
		\includegraphics[height=0.25\textheight,width=0.3\textwidth]
		{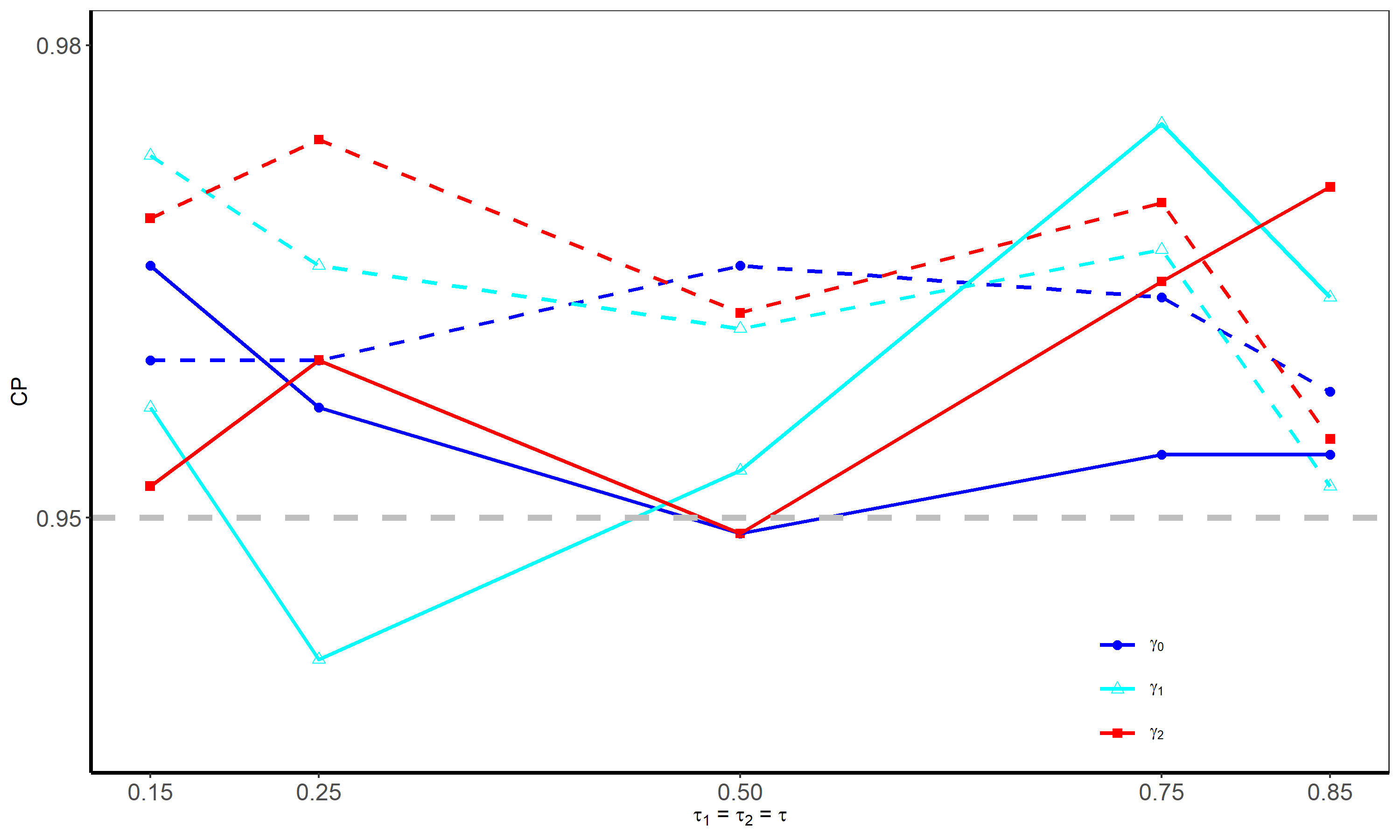}\\
		{\small (a) Bias } & 
		{\small (b) ESE/SD} & 
		{\small (a) CP } 
	\end{tabular}
	\caption{Simulation results for estimated $\bgamma(\btau)$ at $\tau_1=\tau_2=\tau$, with $n=200$ (dashed lines) and $400$ (solid lines).  Bias: empirical bias; ESE/SD: ration of estimated standard error and empirical standard deviation; CP: 95\% coverage probability. }
	\label{fig:sim-est}
\end{figure}

Based on 1000 replications with 
$n=200,400$, 
Figure~\ref{fig:sim-est} presents 
the empirical bias (``Bias"), 
the ratio ESE/SD of the empirical standard deviation (``SD")
and  
the average estimated standard error (``ESE"),  
and the 95\% coverage probability (``CP") 
of $\widehat{\bgamma}(\tau_1,\tau_2)$ at $\tau_1=\tau_2=\tau\in \{0.15,0.25,0.50,0.75,0.85\}$ 
for different scenarios, 
see more numerical results in  Table~B.1 
in Supplementary material.   
We can observe that 
the proposed estimator is asymptotic unbiased, 
the ESE's are close to the SD's, 
and 
the coverage probabilities are close to the nominal level 95\%. 
These observations 
demonstrate that the proposed  estimator has  satisfactory results for all scenarios. 

\subsection{Power of test}
\label{ss:test}

In this section, we conduct simulation studies to assess the performance of the proposed testing procedures.  
Without loss of generality, we restrict attention to the diagonal line association $\tau_1=\tau_2=\tau$, where $\tau\in [t_L, t_U]\subset (0,1)$. 
More specifically, we evaluate the empirical rejection rate of the null hypotheses 
for 
$H_{01}^{(k)}: 
\gamma^{(k)} (\tau)=0$ 
for all $\tau\in [t_L, t_U]$, 
and $H_{02}^{(k)}: 
\gamma^{(k)} (\tau)=c_k$ 
for all $\tau
\in [t_L, t_U]$, where $k=0,1,2$. 
The significance level is set as 0.05 and $[t_L, t_U]=[0.2,0.8]$.

Table~\ref{tab:test} summarizes the empirical rejection percentages of the proposed test procedures in testing $H_{01}^{(k)}$ and $H_{02}^{(k)}$ at significance levels $\alpha= 5\%$. 
Some observations can be drawn from Table~\ref{tab:test}. 
First, 
although slightly conservative, 
the proposed tests have reasonable empirical sizes. 
As expected, as the sample size $n$ increases from $200$ to $400$, the empirical sizes approach closer to the nominal levels. 
Second, all three test statistics have high power in detecting the effect of $Z^{(2)}$ in scenario S2, whereas only $\mathcal{T}_{1,KS}^{(k)}$ and 
$\mathcal{T}_{1,CvM}^{(k)}$ capture the effect of $Z^{(2)}$ in scenario S3. 
This is because $T_{1,W}^{(k)}$ can detect covariate effects on the strength of association but is not sensitive to covariate  effects on association patterns. 
Third, 
the proposed constancy test procedures have reasonably good power in detecting the varying trends in the covariate effects across $\tau$ in scenarios S2 and S3. 

\begin{table}
	\caption{Empirical rejection percentages of significance and constancy tests over 1000 Monte Carlo replications with $B=1000$ bootstrap replications  at significance level $\alpha=5\%$.}
	\label{tab:test}
	\centering
	{\footnotesize
		\resizebox{1\linewidth}{!}{%
			\def\arraystretch{0.75}\tabcolsep=12pt
			\begin{tabular}{cccrrrcrrr}
				\hline
				& & & \multicolumn{3}{c}{$n=200$} & & \multicolumn{3}{c}{$n=400$} \\
				\cline{4-6} \noalign{} \cline{8-10} \noalign{}
				Scenario & Hypothesis & Test & $k=0$ & $k=1$ & $k=2$ & & $k=0$ & $k=1$ & $k=2$ \\ 		
				S1  & $H_{01}^{(k)}$ & $\mathcal{T}_{1,W}^{(k)}$ & 
				0.916 & 0.029 & 0.128 & &  0.996 & 0.040 & 0.184 \\ 
				& & $\mathcal{T}_{1,KS}^{(k)}$ & 
				0.768 & 0.033 & 0.090 & &  0.988 & 0.031 & 0.129 \\ 
				& & $\mathcal{T}_{1,CvM}^{(k)}$& 
				0.895 & 0.023 & 0.115 & &  0.996 & 0.027 & 0.165 \\ 
				&  $H_{02}^{(k)}$ & $\mathcal{T}_{2,KS}^{(k)}$ & 
				0.059 & 0.066 & 0.059 & &  0.045 & 0.043 & 0.046 \\ 
				&  & $\mathcal{T}_{2,CvM}^{(k)}$ & 
				0.047 & 0.040 & 0.039 & &  0.048 & 0.054 & 0.044 \\ 
				S2  &  $H_{01}^{(k)}$ & $\mathcal{T}_{1,W}^{(k)}$ & 
				0.045 & 0.043 & 0.799 &  & 0.053 & 0.045 & 0.978 \\ 
				&   & $\mathcal{T}_{1,KS}^{(k)}$ & 
				0.046 & 0.031 & 0.651 &  & 0.059 & 0.048 & 0.943 \\
				&   & $\mathcal{T}_{1,CvM}^{(k)}$ & 
				0.040 & 0.035 & 0.754 &  & 0.051 & 0.041 & 0.976 \\ 
				&  $H_{02}^{(k)}$ & $\mathcal{T}_{2,KS}^{(k)}$ & 
				0.047 & 0.066 & 0.060 &  & 0.052 & 0.061 & 0.102 \\ 
				&  & $\mathcal{T}_{2,CvM}^{(k)}$ & 
				0.034 & 0.045 & 0.056 &  & 0.041 & 0.043 & 0.118 \\
				S3 & $H_{01}^{(k)}$ & $\mathcal{T}_{1,W}^{(k)}$ & 
				0.965 & 0.033 & 0.024 &  & 0.999 & 0.036 & 0.033 \\ 
				&  & $\mathcal{T}_{1,KS}^{(k)}$ & 
				0.897 & 0.040 & 0.114 &  & 0.999 & 0.042 & 0.307 \\ 
				&  & $\mathcal{T}_{1,CvM}^{(k)}$ & 
				0.960 & 0.035 & 0.084 &  & 0.999 & 0.030 & 0.218 \\ 
				&  $H_{02}^{(k)}$ & $\mathcal{T}_{2,KS}^{(k)}$ & 
				0.210 & 0.068 & 0.209 &  & 0.386 & 0.063 & 0.420 \\ 
				&  & $\mathcal{T}_{2,CvM}^{(k)}$ & 
				0.216 & 0.054 & 0.251 &  & 0.405 & 0.055 & 0.480 \\ 
				\hline
			\end{tabular}%
		}%
	}
\end{table}

\section{Empirical Illustrations}
	\label{s:app}
	
	In this section, 
	we use the proposed method to two empirical applications: 
	one is premature twins data to measure association  between the birth weights, and another is ChIP-seq data analysis for measuring
	reproducibility of high-throughput experiments.

	\subsection{Premature twins data to measure association}
	\label{ss:twins}
	
	Firstly, 
	we use the proposed qcc regression model to analyze the premature twins data~\footnote{The data is openly available at 
		\url{
			https://ysph.yale.edu/c2s2/software/twin-analysis/real-data-set/.}
	}  in \cite{feng2009analysis}.
	All twins were born between January 1, 1994, and December 31, 2004, at three medical centers: Karolinska Institute,
	University of Connecticut, and Yale University. 
	The data included those who survived beyond 36 weeks postmenstrual age. 
	The zygosity of each twin pair was determined
	by histopathologic examination of the placenta with additional confirmation using gender concordance or discordance. 
	Total 63 pairs of monozygotic (MZ) and 137 dizygotic (DZ) twins were identified. 
	Birth
	weight (BW), 
	and gestation age (GA),
	were also available for each infant.

\begin{figure}
		\centering
		\begin{tabular}{cc}
			\includegraphics[height=0.25\textheight,width=0.45\textwidth]{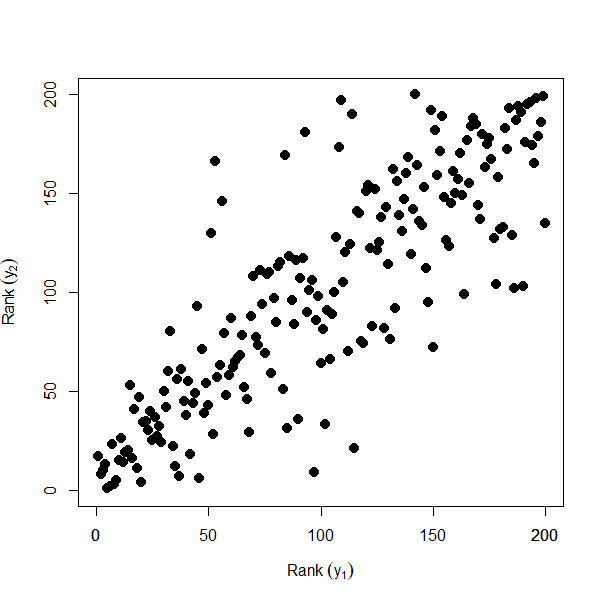} &
			\includegraphics[height=0.25\textheight,width=0.45\textwidth]{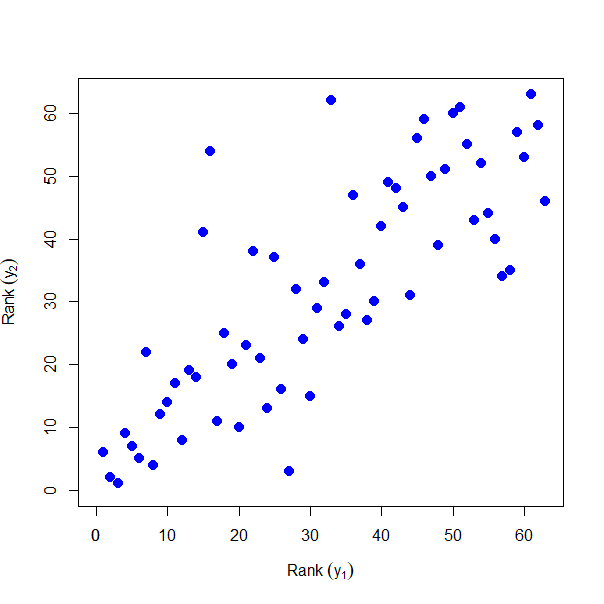} \\
			{\small (a)  All twins (Kendall's tau = 0.659)} & 
			{\small (b) MZ twins (Kendall's tau = 0.620)} \\
			\includegraphics[height=0.25\textheight,width=0.45\textwidth]{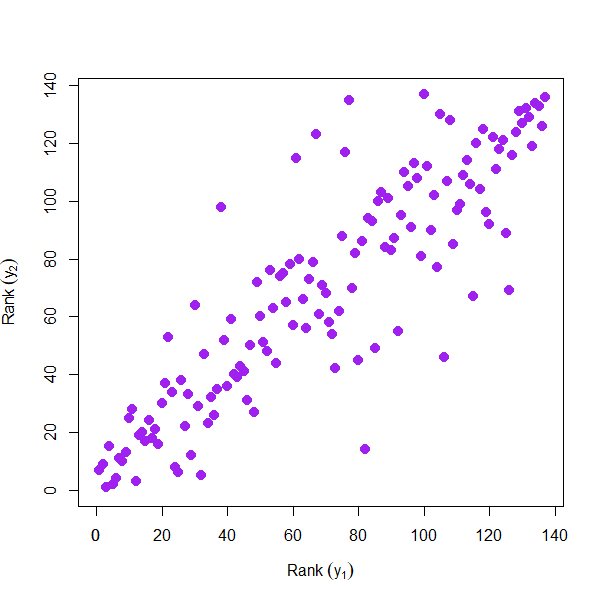} &
			\includegraphics[height=0.25\textheight,width=0.45\textwidth]{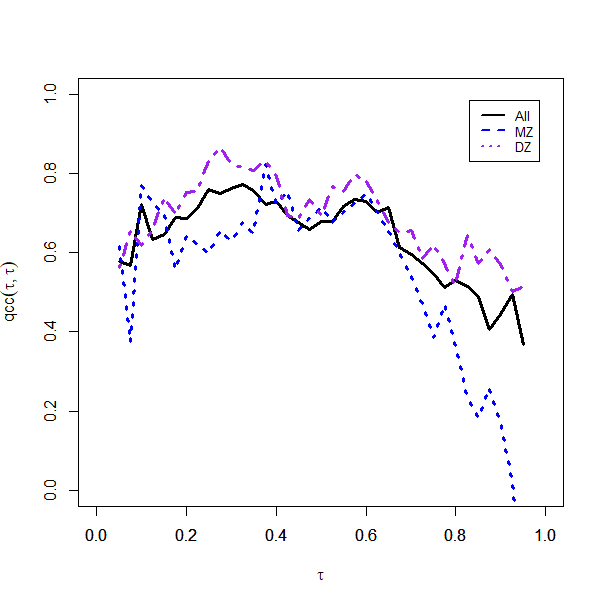} \\
			{\small (c) DZ twins (Kendall's tau = 0.706) } &
			{\small (d) Estimated $\operatorname{qcc}(\tau,\tau)$ curve }
		\end{tabular}
		\caption{Rank scatterplot and estimated $\operatorname{qcc}(\tau,\tau)$ curve of the premature twins data. }
		\label{fig:twins-rank}
	\end{figure}

	The main purpose of this paper is to assess whether and how different covariates explain the association between the birth weights. 
	Figure~\ref{fig:twins-rank} shows the rank scatterplots of the birth weights $(Y_1, Y_2)$ for the whole sample, and two groups (MZ and DZ) of twins. 
	Visually, there is slightly difference in dependence between the birth weights for the MZ twins and DZ twins, 
	whose Kendall's tau coefficients are 0.620 and 0.706, respectively.  
	Particularly,  
	the DZ twins have higher $\operatorname{qcc}(\tau,\tau)$ than the MZ twins on the upper quantiles (e.g., $\tau>0.65$ in Figure~\ref{fig:twins-rank}(d)).  
	It can be observed that genetic factor likely plays an important role for the dependence in birth weights between the twins. 
	
	First, we fit the  conditional quantile correlation $\operatorname{qcc}(\btau|\bZ)$ of twins' birth weight 
	$(Y_1,Y_2)$ by the quantile correlation regression model for $\btau=(\tau_1,\tau_2)$,
	\[
	\phi_F\left(\operatorname{qcc}(\btau|\bZ)\right)
	=\gamma_0(\btau) + \gamma_1(\btau) Z_1 + \gamma_2(\btau) Z_2,
	\]
	where $Z_1$ is the gestational age, 
	$Z_2$ is the zygosity whether the twins are DZ twins (1: DZ twins, 0: MZ twins). 
	As aforementioned, we model covariate effects on the marginal distribution by the linear quantile regression, $Q_{\tau_j}(Y_j|\bZ)=\bZ^\top\bbeta_j(\tau_j)$. 
	For illustration purpose, 
	we focus on equal marginal quantiles $\tau_1=\tau_2=\tau\in[0.2,0.8]$.

	\begin{figure}
		\centering
		\includegraphics[height=0.25\textheight,width=1\textwidth]{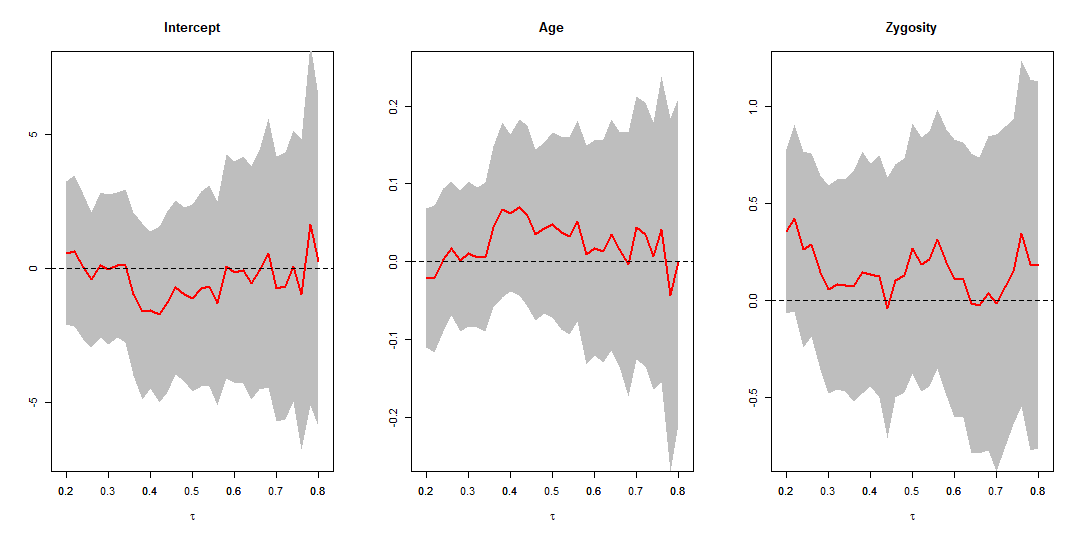}
		\caption{  Estimated regression coefficients $\widehat{\bgamma}(\tau)$ of the premature twins data. Point estimates (red solid line) and their 95\% Wald-type confidence intervals (grey area).}
		\label{fig:twins-est}
	\end{figure}

	Figure~\ref{fig:twins-est} displays the estimated regression coefficients $\widehat{\bgamma}(\tau)$ and their 95\% Wald-type confidence intervals of the premature twins data. 
	One can observe that both intercept and the coefficient for gestation age are not significant. This suggests gestation age do not explain much of the association between the twins' birth weights. 
	However, the conditional association is stronger between DZ twins than between MZ twins, 
	as the coefficient for zygosity is almost greater than zero. 
	A possible explainantion of this interesting pattern is that the MZ twins often sharing one placenta, 
	may cause unbalanced birth weights due to limited nutrition and oxygen~\citep{liu2019flexible}.

	Furthermore, we evaluate the significance and constancy of $\bgamma(\tau)$ jointly for $\tau\in[0.2, 0.8]$. 
	Table~\ref{tab:twins-test} tabulates the testing results with $p$-values  for the premature twins data in different $\mathcal{T}$'s with $B=1000$ bootstrap replications. We can draw the conclusions as follows. 
	First, the tests of no effect for $H_{01}^{(k)}$  in $\mathcal{T}=[0.2,0.4],[0.4, 0.6], [0.6, 0.8], [0.2, 0.8]$ provide $p$-values almost greater than 10\%  for the intercept term, age and zygosity. 
	Second, for the tests of constant effect for $H_{02}^{(k)}$, 	
	all $p$-values almost greater than 10\% for the intercept term, age and zygosity in different $\mathcal{T}$'s, 
	except that $\mathcal{T}_{2,KS}^{(k)}$ in  $\mathcal{T}= [0.2, 0.4]$ provides $p$-values 0.037 and 0.046 for the age and zygosity, respectively, and  $\mathcal{T}_{2,KS}^{(k)}$ provides $p$-value 0.071 for zygosity in  $\mathcal{T}= [0.4, 0.6]$. 
	These results coincide with the results in Figures~\ref{fig:twins-rank}	and~\ref{fig:twins-est}, 
	and also imply the significant zygosity effect on the association of twins' birth weights at lower quantiles.

		\begin{table}
		\caption{Significance and constancy test results ($p$-values) for the premature twins data in different $\mathcal{T}$'s by $B=1000$ bootstrap replications.  
		}
		\label{tab:twins-test}
		\centering
		\begin{footnotesize}
			\resizebox{1\linewidth}{!}{
				{\def\arraystretch{0.8}\tabcolsep=15pt
					\begin{tabular}{cccccc}
						\hline
						$\mathcal{T}$ & Hypothesis &  Test & Intercept & Age & Zygosity \\ 
						\hline
						[0.2, 0.4]  & $H_{01}^{(k)}$ & $\mathcal{T}_{1,W}^{(k)}$ 
						& 0.842 & 0.727 & 0.432
						\\ 
						& & $\mathcal{T}_{1,KS}^{(k)}$
						& 0.751 & 0.625 & 0.169
						\\ 
						& & $\mathcal{T}_{1,CvM}^{(k)}$ 
						& 0.861 & 0.752 & 0.377
						\\ 
						& $H_{02}^{(k)}$ & $\mathcal{T}_{2,KS}^{(k)}$ 
						& 0.119 & 0.037 & 0.046
						\\ 
						& & $\mathcal{T}_{2,CvM}^{(k)}$ 
						& 0.713 & 0.500 & 0.366 
						\\ 
						
						[0.4, 0.6] & $H_{01}^{(k)}$ & $\mathcal{T}_{1,W}^{(k)}$ 
						& 0.595 & 0.470 & 0.594 
						\\ 
						& & $\mathcal{T}_{1,KS}^{(k)}$
						& 0.522 & 0.392 & 0.754
						\\ 
						& & $\mathcal{T}_{1,CvM}^{(k)}$ 
						& 0.592 & 0.445 & 0.649
						\\ 
						& $H_{02}^{(k)}$ & $\mathcal{T}_{2,KS}^{(k)}$ 
						& 0.416 & 0.237 & 0.071
						\\ 
						& & $\mathcal{T}_{2,CvM}^{(k)}$ 
						& 0.957 & 0.901 & 0.707
						\\ 
						
						[0.6, 0.8] & $H_{01}^{(k)}$ & $\mathcal{T}_{1,W}^{(k)}$ 
						& 0.987 & 0.880 & 0.820 
						\\ 
						& & $\mathcal{T}_{1,KS}^{(k)}$
						& 0.983 & 0.994 & 0.880
						
						\\ 
						& & $\mathcal{T}_{1,CvM}^{(k)}$ 
						&  0.994 & 0.973 & 0.945 
						\\ 
						& $H_{02}^{(k)}$ & $\mathcal{T}_{2,KS}^{(k)}$ 
						& 0.921 & 0.869 & 0.197 
						\\ 
						& & $\mathcal{T}_{2,CvM}^{(k)}$ 
						& 0.994 & 0.996 & 0.958
						\\ 
						
						[0.2, 0.8] & $H_{01}^{(k)}$ & $\mathcal{T}_{1,W}^{(k)}$ 
						& 0.786 & 0.642 & 0.586
						\\ 
						& & $\mathcal{T}_{1,KS}^{(k)}$ 
						&  0.910 & 0.820 & 0.454
						\\ 
						& & $\mathcal{T}_{1,CvM}^{(k)}$ 
						&  0.919 & 0.786 & 0.746  
						\\ 
						& $H_{02}^{(k)}$ & $\mathcal{T}_{2,KS}^{(k)}$ 
						&  0.533 & 0.446 & 0.948 
						\\ 
						& & $\mathcal{T}_{2,CvM}^{(k)}$ 
						&  0.976 & 0.949 & 1.000
						\\ 
						\hline
					\end{tabular}
				}
			}
		\end{footnotesize}
	\end{table}

		\subsection{ChIP-seq data analysis to measure reproducibility} 
		\label{ss:chip}
		
		In large-scale omics studies, 
		high-throughput biological experiments are essential tools for identifying biologically interesting candidates 
		with significant associations for further analyses. 
		To measure the reproducibility of the experimental
		outcomes $(Y_{i1},Y_{i2})$, 
		\citet{li2011measuring} proposed 
		the correspondence curve 
		by plotting $(\tau, \widehat{\Psi}_n(\tau))$, 
		where 
		$\widehat{\Psi}_n(\tau)
		=n^{-1}\sum\limits_{i=1}^n
		\left\{Y_{i1}\leq \widehat{Q}_{\tau}(Y_1), Y_{i2}\leq \widehat{Q}_{\tau}(Y_2)\right\}$, which is exactly the sample estimator of $C(\tau,\tau)$.  
		See also \cite{li2018regression} and \citet{zhang2023segmented} for regression frameworks on the correspondence curve. 
		Basically,  the correspondence curve can measure reproducibility 
		by assessing rank consistency of significant candidates between two replicates at a sequence of significance thresholds.   
		Observe that there is a linear transformation between $\widehat{\Psi}_n(\tau)$ and $\widehat{\operatorname{qcc}}(\tau,\tau)$ in \eqref{eqn:qcc-sample}, 
		so it is natural to use $\widehat{\operatorname{qcc}}(\tau,\tau)$ in the case of interest for reproducibility assessment.

		We use the proposed  method to analyze a ChIP-seq dataset in \citet{chen2012systematic} for measuring reproducibility of high-throughput experiments.\footnote{The data is openly available at 
			\url{
				https://github.com/qunhualilab/segCCR.}
		}
		In this study, ChIP-seq experiments for the transcription factor CTCF were performed on two biological replicates,  
		and the CTCF binding sites were identified using the peak calling algorithm useq. 
		The data consisted of 120 million (M) reads, 
		which were generated from Drosophila S2 cells for the site-specific transcription factor Suppressor of Hairy-wing. 
		The deeply sequenced dataset could provide an ideal tool for studying the impact of sequencing depth in ChIP-seq experiments. 
		To assess the impact of sequencing depth, \citet{chen2012systematic} 
		randomly sampled reads from the entire dataset
		to generate subsamples with 0.45, 0.90, 2.70, 5.40 and 16.20M reads, 
		identifing 1335, 2198, 3813, 4631 and
		5499 binding sites  on both replicates, respectively. 
		
		To characterize the effect of sequencing depth on the reproducibility of  two biological replicates  $(Y_{1},Y_{2})$,  
		we fit the conditional quantile correlation $\operatorname{qcc}(\btau|\bZ)$ 
		at $\btau=(\tau_1,\tau_2)$ by   
		\[
		\phi_F(\operatorname{qcc}(\btau|\bZ))
		=
		\gamma_0(\btau) + \gamma_1(\btau) Z_1 + \gamma_2(\btau) Z_2 + \gamma_3(\btau) Z_3 + \gamma_4(\btau) Z_4, 
		\]
		where 0.45M is used as the reference, 
		and 
		$Z_1,Z_2,Z_3, Z_4$  are the dummy variables for 0.90, 2.70, 5.40 and 16.20M, respectively. 
		For simplicity, we take $\tau_1 =\tau_2=\tau\in[0.05,0.95]$.

\begin{figure}
			\centering
			\begin{subfigure}{1\textwidth}
				\includegraphics[height=0.25\textheight,width=1\textwidth]{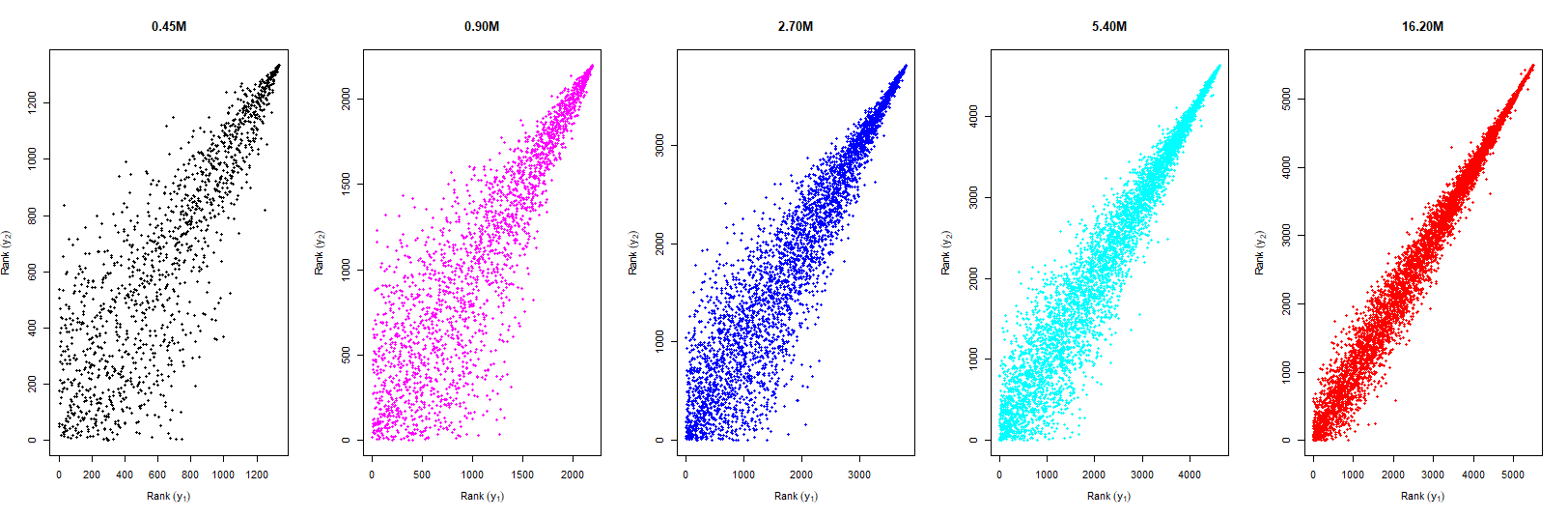}
				\caption{Rank scatterplot}
			\end{subfigure}
			\hspace{1em}
			\begin{subfigure}{1\textwidth}
				\includegraphics[height=0.25\textheight,width=1\textwidth]{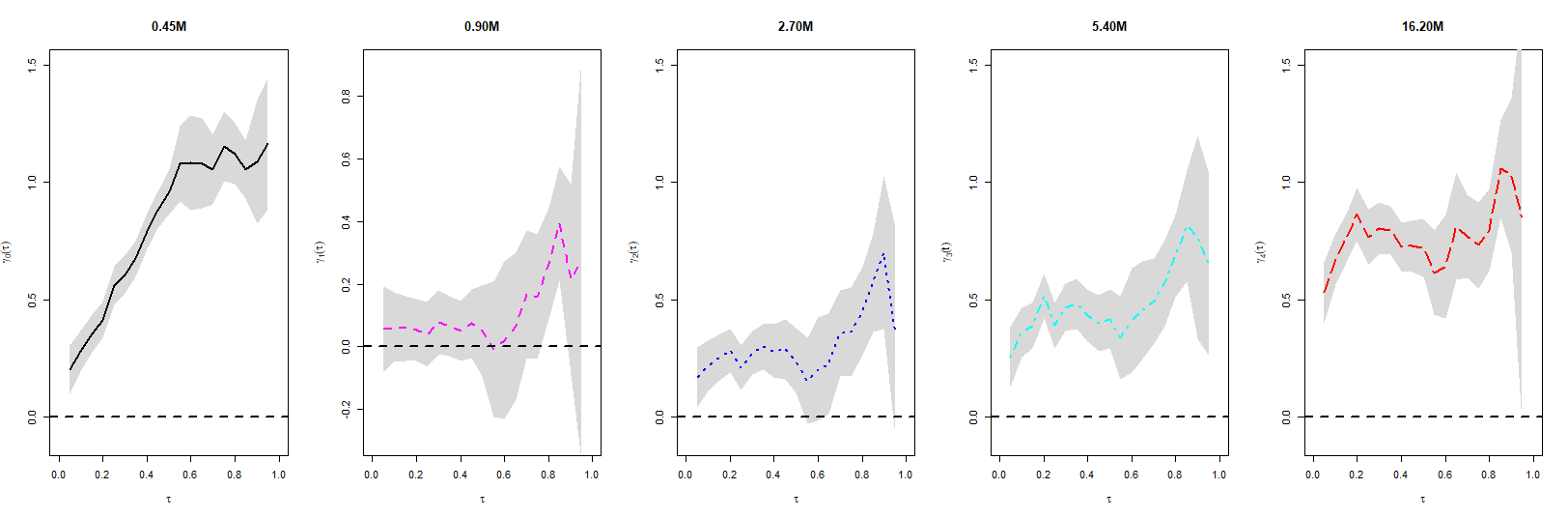}
				\caption{Estimated regression coefficients $\widehat{\bgamma}(\tau)$ and their 95\% confidence intervals}
			\end{subfigure}
			\hspace{1em}
			\begin{subfigure}{0.45\textwidth}	
				\includegraphics[width=2.5in]{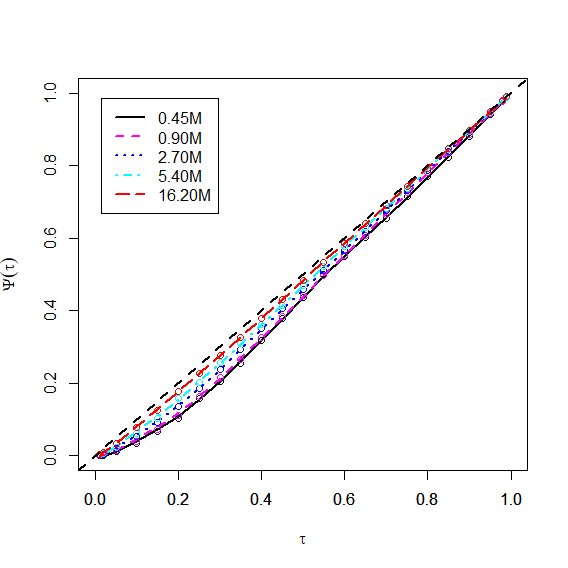}
				\caption{Estimated correspondence curve $\Psi_n(\tau)$}
			\end{subfigure}	
			\hspace{2em}
			\begin{subfigure}{0.45\textwidth}	
				\includegraphics[width=2.5in]{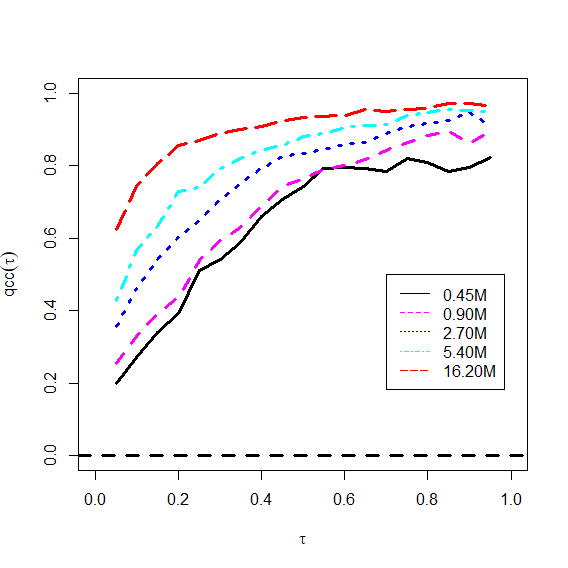}
				\caption{Estimated $\operatorname{qcc}(\tau,\tau)$ curve}
			\end{subfigure}	
			\caption{Rank scatterplot, 
				estimated regression coefficients, 
				correspondence curve $\Psi_n(\tau)$ and $\operatorname{qcc}(\tau,\tau)$ curve of the ChIP-seq data. }
			\label{fig:useq}
		\end{figure}

		As shown in Figure~\ref{fig:useq}(b), 
		the initial increase of sequencing depth
		from 0.45M to 0.90M 
		could result in not significant improvement for reproducibility,  
		because $\gamma_1(\tau)$ is not significant at each quantile level. 
		Once sequencing depth reaches 2.70M, however, 
		reproducibility is significantly improved over 0.45M. 
		This implies that 2.70M may be sufficient for reproducibly identifying strong binding sites. 
		In addition, 
		observing the estimated $\operatorname{qcc}(\tau,\tau)$ curve in Figure~\ref{fig:useq}(d), 
		the reproducibility at each sequencing depth increases with $\tau$.  
		This trend reflects the transition from a less reproducible group to a highly reproducible group, signaling the change from real binding sites to spurious sites on the ranked candidate lists.  
		This also confirms the rank scatterplot and correspondence curve in  Figure~\ref{fig:useq}(a)--(b).  
		Lastly,  the estimated $\Psi_n(\tau)$ curves in Figure~\ref{fig:useq}(c) are visually indistinguishable, 
		but the estimated $\operatorname{qcc}(\tau,\tau)$ curves reveal the differences graphically in different sequencing depths.  
		This illustrates that qcc can serve as an efficient and visual tool to measure reproducibility.

\section{Conclusion}
\label{s:con}

In this paper, 
we have proposed a new local dependence measure, the quantile correlation coefficient, to investigate the  dependence structure of bivariate outcome data. 
Such local dependence measure indexed by the quantiles of the
bivariate outcomes can capture the degree of tail dependent association well. 
Further, we have developed a regression framework for directly modeling covariate effects  on the quantile correlation coefficient.  
The proposed conditional dependence measure has an intuitive interpretation on local dependence strength and facilitates a comprehensive view of dependence structure. 
The proposed estimator demonstrates good finite-sample performances in simulation studies and real data analysis. 
We believe that the proposed method would be very useful for discerning the underlying dependence structure in many applications with bivariate outcomes.

There are several directions for potential future work to enhance the flexibility and robustness of the proposed  framework.  
First, it would be interesting to consider the theoretical results 
when machine learning 
used in place of linear models for the conditional marginal quantile estimators. 
Another promising direction is to extend the proposed framework in regular bivariate data to other complicated data, 
such as survival data with censored outcomes and competing risks.  
We leave these possible interesting topics for future work.

\section*{Disclosure Statement}

The authors report there are no competing interests to declare.

\section*{SUPPLEMENTARY MATERIAL}

The supplementary material includes: 
(i) proofs of the theoretical results; 
(ii) additional results for the simulation studies.

\begin{spacing}{1.7}
	\bibliographystyle{apalike}
	\bibliography{qor-reference}
\end{spacing}



\end{document}